\documentclass[11pt]{article}
\pdfoutput=1
\usepackage{jheppub}
\usepackage{amsmath}
\usepackage{bm}
\usepackage{slashed}
\usepackage{graphicx}
\usepackage{mathrsfs}
\usepackage{enumerate}
\usepackage{stmaryrd}
\usepackage{ytableau}

\newlength{\dummysp}
\newcommand{\beq}{\begin{eqnarray}}
\newcommand{\eeq}{\end{eqnarray}}

\newcommand{\gappeq}{\mathrel{\rlap {\raise.5ex\hbox{$>$}}
{\lower.5ex\hbox{$\sim$}}}}
\newcommand{\lappeq}{\mathrel{\rlap{\raise.5ex\hbox{$<$}}
{\lower.5ex\hbox{$\sim$}}}}

\newcommand{\ben}{\begin{enumerate}}
\newcommand{\een}{\end{enumerate}}

\newcommand{\bit}{\begin{itemize}}
\newcommand{\eit}{\end{itemize}}

\def\[{\left [}
\def\]{\right ]}
\def\({\left (}
\def\){\right )}

\title{\LARGE Large-scale messengers from arbitrary spin fields}

\date{\today}

\author{Mohamed M. Anber} 
\affiliation{Department of Physics, Lewis $\&$ Clark College, Portland, OR 97219, USA}
\emailAdd{manber@lclark.edu}

\abstract{We show that nonperturbative production of arbitrary spin fields from vacuum will accompany the generation of non-vanishing macroscopic energy-momentum tensor correlators. This argument is based on the general causal field formalism, which gives a manifestly covariant description of higher spin particles without any reference to gauge redundancy. Our findings are direct consequence of the Poincar\'e covariance and anlayticity of the Green's functions and independent of any detailed particle physics model. Further, we discuss the idea that any mechanism causing imbalance between the on-shell production of left- and right-handed fields leads to a helical structure in the energy momentum correlators and violation of the macroscopic parity symmetry. We check our method for fields with spin $\frac{1}{2}$ and show that it correctly reproduces previous results. However, the formalism suffers from pathologies related to non-localities that appear for massless particles with spin $\geq 1$ in flat space.  We discuss the origin of these pathologies and the relevance of our findings to cosmology.   
}

\begin{document}
\maketitle

\section{Introduction}
\label{sec:introduction}

Higher spin fields is among the most challenging topics in theoretical physics. Irrespective of the theoretical difficulties, there are a few examples where higher spin fields naturally arise in physical contexts. For example, hadronic resonances have masses of order the strong scale and spins $s \geq 1/2$. String theory, in addition, has a spectrum of higher spin fields with masses above the Planck scale.  QCD excitations can be easily produced in terrestrial high energy experiments. The production of particles with masses larger than a few TeV, however, is infeasible given today's limited technology. Fortunately enough, massive higher spin particles could be produced in the early Universe, e.g., during inflation, electroweak, or QCD phase transitions. This will happen provided there is a mechanism that leads to the production of these particle, e.g., parametric resonance \cite{Shtanov:1994ce}. Generally, the non-perturbative production of particles from vacuum  will be accompanied by the generation of gravitational waves (GW) \cite{Anber:2016yqr,Sorbo:2011rz}. Despite the fact that  heavy particles will immediately decay once they are produces, GW accompanying their production might be detected today as a signal from the early epoch of the Universe. One hopes that these GW will have distinct features that distinguish them among other early Universe signals.\footnote{Non-Gaussianities from higher spin fields was discussed in \cite{Arkani-Hamed:2015bza}. See also \cite{Baumann:2017jvh,Franciolini:2017ktv} for recent discussions of the imprints of higher spin fields on cosmological perturbations.}

A systematic study of higher spins started as early as quantum field theory itself. However, this subject remains a largely unexplored territory until today.  In 1939, Fierz and Pauli were the first to write a system of Lorentz covariant equations that respect unitarity and describe the motion of higher spin fields in flat background \cite{Fierz:1939ix}. In their studies, they noticed that turning on interactions among higher spin particles results in inconsistencies in the field theories describing them. This led them to suggest that a Lagrangian formulation of higher spin fields might evade these difficulties and renders the theory consistent. It was not until the mid 1970s when  Singh, Hagen,  Fronsdal, and Fang, completed the program of Feirz and Pauli by writing a Lagrangian formulation of fields with arbitrary spins \cite{Fang:1978wz,Fronsdal:1978vb,Fronsdal:1978rb,Singh:1974qz,Singh:1974rc}. This formulation (known as Fronsdal formulation) extends the concept of gauge field theory of electromagnetism and gravity to an arbitrary spin and demands the introduction of auxiliary fields that are used to eliminate the spurious degrees of freedom. This Lagrangian formulation, however, was limited to free field theories.   In the meanwhile, many no-go theorems appeared in the period from 1960s to 1980s that forbid the minimal coupling between higher spin fields and electromagnetism or gravity, see \cite{Vasiliev:1999ba,Bouatta:2004kk,Rahman:2015pzl} for reviews. In addition, Weinberg and Witten proved that there cannot be a consistent massless field theory with spin $s\geq 2$ that has a gauge invariant and conserved energy-momentum tensor \cite{Weinberg:1980kq}. However, starting from 1980 many yes-go examples of interacting higher spin fields appeared. In all these examples non-minimal coupling, e.g., \cite{Porrati:1993in}, and/or formulating the theory on a curved background, e.g., \cite{Fradkin:1987ks}, was a necessary ingredient for a consistent description of higher spin particles.

 Independently, Bragmann and Wigner \cite{Bargmann:1948ck} and Weinberg \cite{Weinberg:1964cn,Weinberg:1964ev} wrote down a free field description of higher spin particles based on the higher dimensional irreducible  representations of Poincar\'e algebra. These fields are known as the {\em general causal fields} \cite{Weinberg:1995mt}. This construction generalizes the Weyl equation, which describes the motion of a spin half particle belonging to $(1/2,0)$ or $(0,1/2)$ representation of the Lorentz algebra, to higher spin fields that transform as $(s,0)$ or $(s,0)$ under the Lorentz group. Alternatively, we can also use the Dirac representation $(s,0)\oplus(0,s)$, which is particularly important in the case of massive higher spin particles.  Here, one does not introduce any gauge redundancy since the fields used in this description are the physical ones. Therefore, no question of inconsistency or unphysical states can arise. The price one pays by working with the physical fields, however, is the lack of a Lagrangian formulation which makes their coupling to a background field, like electromagnetism or gravity, a tricky business. 

In this work we examine the generation of large scale signals from the nonperturbative production of general causal fields. In particular, we seek a model-independent setup that can provide simplified answers about the generation of GW that accompany the parametric-resonance production of higher spin fields. A precise account of the details of such scenario is a daunting task, specially that coupling higher spin fields to gravity (not to mention their coupling to other fields, which is necessary to provide the parametric resonance mechanism) is still an open question. In our setup, however, we do not need to know about the specific particle-physics model that explains the generation of higher spin fields from vacuum. We just assume that these fields are somehow produced, and therefore, we only need to compute their on-shell energy-momentum tensor correlators. This should be translated into large scale GW signals via multiplying by appropriate transfer, i.e. Green's, functions.  The normal procedure to compute the energy-momentum tensor correlator, which accompanies the production of particles in the presence of an external field or time varying background, e.g., in a cosmological context, is to start from a Lagrangian, vary it with respect to the metric tensor to obtain the energy-momentum tensor,  and then use the canonical quantization to write down the fields in terms of creations and annihilation operators. Next, we evolve these operators via the Bogoliubov transformation to finally obtain an expression of the energy-momentum tensor and its correlators. The absence of a Lagrangian formulation for the general causal fields, however, makes it necessary to find an alternative route that enables us tackle the problem indirectly. Motivated by the effective action of $s=\frac{1}{2}$ field along with the principle of Poincar\'e covariance, we postulate a definition of the energy-momentum tensor of general causal fields in a flat or curved background. Then, we use the analytic properties of Green's function to show that the nonperturbative production of higher spin fields from vacuum will accompany the emergence of non-vanishing energy-momentum tensor correlators. 

As a check on our new formalism, we compute the energy-momentum tensor correlator of $s=1/2$ fields in flat and FRW backgrounds reproducing  previous results that were obtained using the canonical formalism. Further, we apply our method to compute the energy-momentum tensor correlators of massive spin particles with spin $s>1/2$ and show that these correlators respect unitarity. This can be shown  by projecting the correlators along helicity-2 eigenbasis, which is a succinct way to directly check that $\langle T_-(k)T_-(-k) \rangle>0$ and $\langle T_+(k)T_+(-k) \rangle>0$, where $T_{\pm}$ are the energy-momentum tensor along the positive (negative) helicity-basis. Interestingly enough, these two correlators may not be equal in the presence of a mechanism that favors one helicity over the other. This can also be envisaged directly from our construction without the need to provide a detailed particle-physics model to explain it. 

However, we show that the production of massless fields with spin $s\geq 1$ in a flat background is pathological within the general causal framework. This is attributed to the fact that the emission of a massless spin $s$ particle vanishes as $|\bm p|^{s-\frac{1}{2}}$, where $|\bm p|$ is its momentum.   Nevertheless, our work is a proof of concept that the production of higher spin fields will accompany the generation of macroscopic energy-momentum tensor correlators, which may also signal the breaking of the macroscopic parity symmetry.  

Our work in organized as follows. In Section \ref{sec:theory} we provide the necessary mathematical background of the general causal fields in both flat and curved backgrounds. In particular,  we review the Lorentz group and Weinberg construction of Green's functions of the general causal fields in flat space. This includes both massless and massive fields. Next, we generalize this construction to curved background using the vierbein formalism. We also provide a review section on spinor calculus. In Section \ref{Path integral, effective action, and energy momentum tensor} we use the effective action of the spin $\frac{1}{2}$ field along with the principle of Poincar\'e covariance as a motivation to postulate a definition of the energy-momentum tensor of higher spin general causal fields. The energy-momentum tensor and its correlators contain vertex functions that we determine in Section \ref{Vertex operators}. In Section \ref{Structure of the energy-momentum correlator} we use the formalism of Sections \ref{sec:theory} to \ref{Vertex operators} along with the  analyticity of Green's functions to show that the production of higher spin fields from vacuum accompanies the emergence of macroscopic energy-momentum tensor correlator. We calculate this correlator for $s=\frac{1}{2}$ and $s=1$ in flat background and also for $s=\frac{1}{2}$ in FRW background. Using the energy-momentum tensor correlator, we review the gravitational waves power spectrum in Section \ref{Gravitational waves}. We conclude in Section \ref{Discussion} by discussing the implications and limitations of our work.

\section{Theory and formulation}
\label{sec:theory}

\subsection{General causal fields}

 This section aims to introduce the formalism and notation used throughout this work. We use the general causal fields (also called the Weinberg-type fields) to describe higher spin particles. This formalism is heavily based on the higher dimensional representations of the Lorentz algebra $so(3,1)$ or its double cover $sl(2,\mathbb C)$ that we will review momentarily. In this work we use the signature $\eta_{ab}=(-,+,+,+)$. The Greek letters are used to denote the curved spacetime coordinates, the latin letters $a,b,c,d$ are used for the flat spacetime coordinates, the latin letters $i,j,k$ label the spatial flat coordinates, while the latin letters $M,N,P,Q$ are used to label the matrix elements of $sl(2,\mathbb C)$. The Lorentz algebra generators ${\cal T}_{ab}$ satisfy the commutation relations \cite{Tung:1985na}
\begin{eqnarray}
\left[{\cal T}_{ab}, {\cal T}_{cd} \right]=i\left(\eta_{cb}{\cal T}_{ad}-\eta_{ca}{\cal T}_{bd}+\eta_{db}{\cal T}_{ca}-\eta_{da}{\cal T}_{cb}\right)\,.
\label{Lorentz algebra CR}
\end{eqnarray}
In particular we have ${\cal T}_{ab}=-{\cal T}_{ba}$ such that
\begin{eqnarray}
{\cal T}_{i0}=K_i\,,\quad \frac{1}{2}\epsilon_{ijk}{\cal T}_{jk}=J_i\,,
\end{eqnarray}
where $K_i$ and $J_i$ are respectively the generators of boosts and $so(3)$ rotations. Then, it is a simple exercise to use (\ref{Lorentz algebra CR}) to show that the combinations
\begin{eqnarray}
{\cal A}_i=\frac{1}{2}\left(J_i+iK_i\right)\,,\quad
{\cal B}_i=\frac{1}{2}\left(J_i-iK_i\right)
\end{eqnarray}
satisfy the commutation relations of two independent copies of $su(2)$ algebra:
\begin{eqnarray}
\left[{\cal A}_i,{\cal A}_j\right]=i\epsilon_{ijk}{\cal A}_i\,,\quad
\left[{\cal B}_i,{\cal B}_j\right]=i\epsilon_{ijk}{\cal B}_i\,,\quad
\left[{\cal A}_i, {\cal B}_j\right]=0\,~\mbox{for all}\,~ i,j\,.
\end{eqnarray}
The two algebras are called $su_{L}(2)$ and $su_{R}(2)$ for $\{{\cal A}_i\}$ and $\{{\cal B}_i\}$ generators, respectively. Let the set of generators $\{{\cal C}_i\}$ denotes either $\{{\cal A}_i\}$ or $\{{\cal B}_i\}$. Then for a representation of spin $s$ we have 
\begin{eqnarray}
\left[{\cal C}_1^{(s)}\pm i {\cal C}_2^{(s)} \right]_{M}^N=\delta_{M,N\pm M}\left[(s\mp N)(s\pm N+1)  \right]^{1/2}\,, \quad \left[{\cal C}_3^{(s)}\right]_{M}^N=M\delta_{MN}\,,
\end{eqnarray}
where $M,N=s,s-1,...,-s+1,-s$ and $s=0,1/2,1,3/2,...$

The complexified generators of  $so(3,1)$ are related to $su(2)_L \oplus su(2)_R$, and hence, any representation of the Lorentz algebra can be designated by a pair  of numbers $(s_1,s_2)$, where $s_{1}$ and $s_2$ are the spins of the representations of $su(2)_L$ and $su(2)_R$, respectively. Thus, the generators ${\cal T}_{ab}$ can be written as the direct sum of two sets of generators transforming under  $su(2)_L$ and $su(2)_R$ such that ${\cal T}_{ab}\equiv I^{su(2)_L}\otimes{\cal T}_{ab}^{su(2)_R}+{\cal T}_{ab}^{su(2)_L}\otimes I^{su(2)_R}$, or explicitly using the matrix elements $\left[{\cal T}_{ab}\right]^{M_1M_2}_{N_1N_2}=\delta^{M_1}_{N_{1}}\left[{\cal T}_{ab}\right]^{M_2}_{N_2}+\left[{\cal T}_{ab}\right]^{M_1}_{N_1}\delta^{M_2}_{N_{2}}$, where $M_1,N_1=s_1,s_1-1,...,-s_1+1,-s_1$ and $M_2,N_2=s_2,s_2-1,...,-s_2+1,-s_2$. 

Under the Lorentz transformation, a general field  transforms in the $s_1$ representation of $su(2)_L$ and $s_2$ representation of $su(2)_R$. As special cases, we consider fields with $s_1=0$ or $s_2=0$ such that: 
\begin{eqnarray}
\begin{array}{cc}
 \bm J=\bm J^{(s)}\,,\bm K=-i \bm J^{(s)} & \mbox{for type}~(s,0)~\mbox{fields}\,,\\\\
\bm J=\bm J^{(s)}\,,\bm K=+i \bm J^{(s)}& \mbox{for type}~(0,s)~\mbox{fields}\,,
\end{array}
\label{left and right fields}
\end{eqnarray}
where a boldface symbol denotes a three dimensional vector.
Such fields transform as $2s+1$ dimensional spinors and are denoted by $\Phi_L$ for $(s,0)$ and $\Phi_R$ for $(0,s)$ such that $\Phi^{(s)}_{L,R}\equiv\left(\Phi_1,\Phi_2,...,\Phi_{2s+1}\right)^T$. They are known as the {\em general causal fields} \cite{Weinberg:1995mt} or Weinberg-type fields \cite{Weinberg:1964cn,Weinberg:1964ev}. Under a general proper Lorentz transformation $\Lambda$ (we can always take the boost to be along the $z$-direction without loss of generality) these fields transform as
\begin{eqnarray}
\nonumber
\Phi^{(s)}_L&\rightarrow& D^{(s)}[\Lambda]\Phi^{(s)}_L=D^{(s)}[R]e^{-i\phi K_3^{(s)}}\Phi_L^{(s)}~\mbox{for type}~(s,0)~\mbox{fields}\,,\\
\Phi^{(s)}_R&\rightarrow& \bar D^{(s)}[\Lambda]\Phi^{(s)}_R=D^{(s)}[R]e^{-i\phi K_3^{(s)}}\Phi_R^{(s)}~\mbox{for type}~(0,s)~\mbox{fields}\,,
\end{eqnarray}
where $\phi$ is the boost parameter (rapidity) along the $z$-direction and $D^{(s)}[R]$ is the $so(3)$ part of the Lorentz transformation matrix. 

Now consider a $(s,0)$ massless field moving along the $z$-direction with its spin directing along the positive $z$-axis: $\Phi^{(s)}_L=\left(0,0,0,...,\underbrace{1}_{2s+1}\right)^T$. Then, it is trivial to show that such field satisfies 
\begin{eqnarray}
J_3^{(s)}\Phi^{(s)}_L=-s\Phi^{(s)}_L\,.
\end{eqnarray}
By applying a general rotation about the $z$-axis we can write the above equation as
\begin{eqnarray}
\bm p \cdot \bm J^{(s)}\Phi^{(s)}_L=-s|\bm p|\Phi^{(s)}_L\,,\quad\mbox{for}~(s,0)~\mbox{fields}\,,
\label{left helical fields}
\end{eqnarray}
where $\bm p$ is the field momentum. Thus, the $(s,0)$ fields describe left-handed particles with helicity $-s$. By the same token, the $(0,s)$ fields satisfy
\begin{eqnarray}
\bm p \cdot \bm J^{(s)}\Phi^{(s)}_R=+s|\bm p|\Phi^{(s)}_R\,,\quad\mbox{for}~(0,s)~\mbox{fields}\,,
\label{right helical fields}
\end{eqnarray}
and they describe right-handed particles with helicity $+s$.  Now, it is not difficult to see that (\ref{left helical fields}) and (\ref{right helical fields}) can be rewritten as first order differential equations:
\begin{eqnarray}
\nonumber
&&\bm J^{(s)}\cdot \nabla \Phi^{(s)}_L-s\partial_t \Phi^{(s)}_L=0\,, \quad \mbox{for}\,\, (s,0)~\mbox{fields}\,,\\
&&\bm J^{(s)}\cdot \nabla \Phi^{(s)}_R+s\partial_t \Phi^{(s)}_R=0\,, \quad \mbox{for}\,\, (0,s)~\mbox{fields}\,.
\label{1st order EOM}
\end{eqnarray} 
Every equation describes the propagation of a free single degree of freedom. In addition, both types of fields satisfy the second order Klein-Gordon equation $\Box \Phi_{L,R}=0$. In fact, Equations (\ref{1st order EOM}) are the Weyl equations for $s=1/2$.  With a bit of work, one can show that the case $s=1$ corresponds to Maxwell's equations for the left and right circularly polarized radiation in free space\footnote{For example, using the assignment $\bm \psi=\bm E-i\bm B$, one can show the equivalence between the first equation in (\ref{Maxwell polarized}) and the first equation in (\ref{1st order EOM}) via the transformation ($\Phi$ denotes the left-handed field):
\begin{eqnarray}
\psi_1=\frac{-i}{\sqrt 2}\left(\Phi_1-\Phi_3\right)\,,\quad \psi_2=\frac{1}{\sqrt 2}\left(\Phi_1+\Phi_3\right)\,,\quad \psi_3=i\Phi_2\,.
\end{eqnarray}
}:
\begin{eqnarray}
\nonumber
\nabla \times \left(\bm E-i\bm B\right)+i\frac{\partial}{\partial t}\left(\bm E-i\bm B\right)&=&0\,, \quad \mbox{for}\,\, (1,0)~\mbox{fields}\,,\\
\nabla \times \left(\bm E+i\bm B\right)-i\frac{\partial}{\partial t}\left(\bm E+i\bm B\right)&=&0\,, \quad \mbox{for}\,\, (0,1)~\mbox{fields}\,.
\label{Maxwell polarized}
\end{eqnarray}
Moreover, applying the divergence operator on the above equations we obtain the Bianchi identity $\nabla \cdot \bm E=\nabla \cdot \bm B=0$.
Since the equations of motion are first order, there are no pathologies associated with the higher spins when we use the general causal construction. Notice also that here we work directly with the field strengths (the physical fields) rather than the potential fields, as is evident from $s=1/2$ and $s=1$ examples. 

The massless general causal  fields are chiral by construction. Thus, under the parity transformation ${\cal P}:\bm r\rightarrow -\bm r$ we have
\begin{eqnarray}
\bm K \stackrel{{\cal P}}{\rightarrow} -\bm K\,,\quad \bm J \stackrel{{\cal P}}{\rightarrow} \bm J\,,
\end{eqnarray}
and hence,
\begin{eqnarray}
\Phi^{(s)}_L \stackrel{{\cal P}}{\rightarrow} \Phi^{(s)}_R \,.
\end{eqnarray}
For massless fields chirality and helicity coincide, and both work as a good Lorentz invariant quantum number. We will also limit our analysis to the left-handed fields $(s,0)$ since the right-handed ones follow the exact same construction. The left handed fields will be denoted by $\Phi$, with no $L$ subscript, when no confusion can arise. 

We can also use the Dirac's representation  $(s,0)\oplus(0,s)$. In this case one combines the left and right fields in a single field $\Psi^{(s)}$ as
\begin{eqnarray}
\Psi^{(s)}=\left[ \begin{array}{c}\Phi_L^{(s)}\\ \Phi_R^{(s)} \end{array} \right]\,,
\label{def Psi field}
\end{eqnarray}
such that  $\Psi^{(s)}$ respects the parity, charge conjugation, and time reversal symmetries. This construction is particularly important when we deal with massive fields. 

\subsection{Quantization of the general causal fields in flat space: massless fields}

After discussing the classical aspects of the general causal fields, now we  turn into their canonical quantization. As we mentioned above,  a free left-handed particle moving along the $z$-direction, with reference momentum $\kappa$, is described by the spinor $\xi^{(s)}=\left(0,0,...,0,\underbrace{1}_{2s+1}\right)^T$. One can obtain the general form of the spinor by applying a general proper Lorentz transformation $\Lambda$ (boost and rotation):
\begin{eqnarray}
\label{transformation of xi}
\xi^{(s)} \rightarrow \xi^{(s)}(\bm p)= D^{(s)}[\Lambda]\xi^{(s)}\equiv D^{(s)}[R({ \bm p})]e^{-i\phi(|\bm p|)K_3}\xi^{(s)}\,,
\end{eqnarray}
where the matrix $D^{(s)}[R({ \bm p})]$ is the rotational part of the Lorentz transformation\footnote{The matrix $D^{(s)}[R( \bm p)]$ is an $so(3)$ rotation that takes the form
\begin{eqnarray}
D^{(s)}[R({ \bm p})]=e^{-i\hat {\bm n}\cdot \bm J^{(s)}\theta}\,,\quad \hat{\bm n}=\frac{\left(-p_y,p_x,0\right)}{\sqrt{p_x^2+p_y^2}}\,,\quad \theta=\cos^{-1}\left(\frac{p_z}{|\bm p|}\right)\,.
\end{eqnarray}}, while $e^{-i\phi(|\bm p|)K_3}$ is the boost. One can show easily that the rapidity $\phi$ is given by $\phi(|\bm p|)=\log \left[|\bm p|/\kappa\right]$, where $\bm p$ is the particle's momentum after applying the boost. 

 Now, the quantization of the general causal fields can be achieved by carrying out the exact same steps used in the quantization of a spin $1/2$ field. One expands the classical field in terms of a complete set of orthonormal states with complex coefficients. Then, we promote these coefficients to creation and annihilation operators. In this work we are interested in the production of on-shell higher spin particles, and hence, all our fields and their Green's functions will be assumed to satisfy the on-shell condition.   The quantum field $\Phi^{(s)}$, then, takes the following canonical form in the Fourier space 
\begin{eqnarray}
\Phi_M^{(s)}(x)=\int \frac{d^3 \bm p}{(2\pi)^3 \sqrt{2p^0}}\left[\frac{|\bm p|}{\kappa}\right]^s\left[a_{\bm p}e^{i  \bm p\cdot \bm x-i p^0t} +b_{\bm p}^\dagger e^{-i \bm p\cdot \bm x+i p^0 t} \right]D_{M,-s}^{(s)}[R({ \bm p})]\,,
\label{fields in terms of a and a dagger}
\end{eqnarray}
where the on-shell condition, $p^0=|\bm p|$, has been assumed and we have used (\ref{transformation of xi}) and (\ref{left and right fields}). The annihilation and creation operators $a_{\bm p}$, $a^\dagger_{\bm p}$,  $b_{\bm p}$, and $b^\dagger_{\bm p}$ satisfy the the (anti)commutation relations  
\begin{eqnarray}
\left[a_{\bm p},a^\dagger_{\bm p'} \right]_{\pm}=(2\pi)^3\delta^3(\bm p-\bm p')\,, \quad \left[b_{\bm p},b^\dagger_{\bm p'} \right]_{\pm}=(2\pi)^3\delta^3(\bm p-\bm p')
\label{commutations of a and b}
\end{eqnarray}
where the commutator (the $-$ sign) is used for integer spin fields and the anti-commutator (the $+$ sign) is used for the half-integer spin fields. Then, it can be easily shown that these fields satisfy the microcausality condition
\begin{eqnarray}
\left[\Phi_{M}^{(s)}(x), \Phi_N^{\dagger(s)} (x')\right]_{\pm}=-\Pi_{MN}(-i\partial)\Delta(x-x')\,,
\label{commutation relation for fields}
\end{eqnarray}
and the matrices $\Pi_{MN}$ are defined by (\ref{general forms of Pi}) below and  $\Delta(x-x')$ is the scalar propagator. In fact, the (anti)commutator relations (\ref{commutation relation for fields}) establish the connection between spin and statistics. It is customly to redefine the normalization of the massless field $\Phi$ by replacing the factor $\kappa^{-s}$ with $2^{s}$, which we will do in the rest of this work. 

Notice that in writing the quantum field $\Phi_M^{(s)}(x)$, which describes a single degree of freedom,  we used the annihilation and creation operators of the physical particles, which are described by the spinor $D_{M,-s}^{(s)}[R({ \bm p})]$. This spared us from the need to introduce any spurious degrees of freedom that could lead to unphysical negative energy states or inconsistencies. 

\subsection*{Structure of the massless Green's functions}

As we show in Section (\ref{Path integral, effective action, and energy momentum tensor}) the Green's functions are indispensable tool to couple general causal fields to a fixed background. The time-ordered Green's function for the left-handed fields is defined as
\begin{eqnarray}
G_{0\,MN}^{(s)}(\bm x,\bm x', t,t')=\langle 0| {\cal T}\Phi^{(s)}_M(\bm x,t)\Phi^{(s)\dagger}_N(\bm x',t')|0\rangle\,,
\end{eqnarray}
where the time-order operator ${\cal T}$ is
\begin{eqnarray}
{\cal T} \Phi^{(s)}_M(\bm x,t)\Phi^{(s)\dagger}_N(\bm x',t')=\left\{\begin{array}{cc}  \Phi^{(s)}_M(\bm x,t)\Phi^{(s)\dagger}_N(\bm x',t') & t>t'\\ (-1)^{2s} \Phi^{(s)\dagger}_N(\bm x',t') \Phi^{(s)}_M(\bm x,t)& t<t' \end{array}\right.\,.
\end{eqnarray}
Now, using (\ref{fields in terms of a and a dagger}) and $a|0\rangle=b|0\rangle=0$ we can write the Green's function in the form
\begin{eqnarray}
\nonumber
G_{0\,MN}^{(s)}(\bm x,\bm x', t,t')=\int \frac{d^3 p}{\left(2\pi\right)^3}e^{i \bm p\cdot \left(\bm x-\bm x'\right)}{\cal G}_{0\,MN}^{(s)}(\bm p, \bm p',t,t')
\label{time ordered Greens function}
\end{eqnarray}
where the momentum-space Green's function is
\begin{eqnarray}
{\cal G}_{0\,MN}^{(s)}(\bm p, \bm p',t,t')=\frac{|\bm p|^{2s}}{2|\bm p|}\left\{ \begin{array}{cc}\widehat{\Pi_{MN}}(p^0,{\bm p})e^{i|\bm p|(t-t')}\,,& t>t'\\ (-1)^{2s} \widehat{\Pi_{MN}}(p^0,-{\bm p})e^{i|\bm p|(t'-t)}
\,, & t<t' \end{array}  \right.\,,
\label{momentum space left handed particles}
\end{eqnarray}
and $\widehat{\Pi_{MN}}(p^0,{\bm p})=D_{M,-s}^{(s)}[R({ \bm p})]D_{N,-s}^{*(s)}[R({ \bm p})]$. The matrices $\widehat{\Pi_{MN}}$ can be put in a nicer form by using the formula (which can easily be proved by induction):
\begin{eqnarray}
\delta_{M,-s}\delta_{N,-s}=\frac{1}{(2s)!}\left[\prod_{\lambda=-s+1}^s\left(\lambda \mathbb I-J_3\right)\right]_{MN}\,,
\end{eqnarray}
and then applying the rotation matrix $D^{s}[R({ \bm p})]$ to find:
\begin{eqnarray}
\widehat{\Pi_{MN}}(p^0,{\bm p})=\frac{1}{(2s)!|\bm p|^{2s}}\left[\prod_{\lambda=-s+1}^s\left(\lambda p^0\mathbb I -\bm p\cdot \bm J\right)\right]_{MN}\,.
\label{Pi in terms of p}
\end{eqnarray}
In Section (\ref{Spinor calculus}) we show that one can write $\widehat{\Pi_{MN}}(p^0,{\bm p})$ in the Lorentz covariant form
\begin{eqnarray}
\widehat{\Pi_{MN}}(p^0,{\bm p})=\frac{(-1)^{2s}}{2^{2s}}t_{MN}^{a_1a_2...a_{2s}}n_{a_1}(\bm p)n_{a_2}(\bm p)...n_{a_{2s}}(\bm p)\,,
\label{Pi in terms of t first equation}
\end{eqnarray} 
where $n_{a}\equiv (1, \frac{\bm p}{|\bm p|})$ are light-like vectors and $t_{MN}^{a_1a_2...a_{2s}}$ are the generalized Pauli matrices as we explain in Section (\ref{Spinor calculus}). 

\subsection{Quantization of the general causal fields in flat space: massive fields}

In the massive case it is more convenient to work with the Dirac representation $(s,0)\oplus(0,s)$, where one combines the left and right fields in a single field $\Psi^{(s)}$ as defined in (\ref{def Psi field}). The field $\Psi^{(s)}$ satisfies the massive Klein-Gordon equation
\begin{eqnarray}
(\Box-m^2)\Psi^{(s)}(x)=0\,.
\end{eqnarray}
In addition, one can show that $\Psi^{(s)}(x)$ satisfies the generalized Dirac's equation
\begin{eqnarray}
\left[-i^{2s}\gamma^{a_1...a_{2s}}\partial_{a_1}...\partial_{a_{2s}}+m^{2s} \right]\Psi^{(s)}(x)=0\,,
\label{generalized dirac equation}
\end{eqnarray}
where the generalized $\gamma$ matrices, $\gamma^{a_1...a_{2s}}$, are given by
\begin{eqnarray}
\gamma^{a_1...a_{2s}}=\left[\begin{array}{cc} 0 & t^{a_1...a_{2s}} \\ \bar t^{a_1...a_{2s}} & 0 \end{array} \right]\,,
\end{eqnarray}  
where the matrices $t$ and $\bar t$ will be introduced in Section (\ref{Spinor calculus}).
Unlike the massless case, where the left- or right- handed fields describe the propagation of a single degree of freedom, massive fields on the other hand  describe $(2s+1)$ massive degrees of freedom for $(s,0)$ or $(0,s)$ fields. This is obvious since the dimension of $SO(3)$ group (which is the little group in the massive case)\footnote{Notice that the little group in the massless case is $SO(2)$, and hence, the Dirac field has only $4$ degrees of freedom.} has dimension $(2s+1)$.  

Next, as usual, we expand the left and right fields in terms of creation and annihilation operators as follows:
\begin{eqnarray}
\nonumber
\Phi_{L\,,M}^{(s)}&=&\int \frac{d^3 p}{(2\pi)^3\sqrt{2\omega (\bm p)}}\sum_{N}\left\{D^{(s)}_{MN}({\bm p})a_{\bm p, N}e^{ip\cdot x}+D^{(s)}_{MN}({\bm p})b^\dagger_{\bm p,N}e^{-i p\cdot x} \right\}\,,\\
\nonumber
\Phi_{R\,,M}^{(s)}&=&\int \frac{d^3 p}{(2\pi)^3\sqrt{2\omega (\bm p)}}\sum_{N}\left\{D^{(s)}_{MN}(-{\bm p})a_{\bm p, N}e^{ip\cdot x}+(-1)^{2s}D^{(s)}_{MN}(-{\bm p})b^\dagger_{\bm p,N}e^{-i p\cdot x} \right\}\,,\\
\label{left and right field expansions in flat space}
\end{eqnarray} 
where $\omega(\bm p)=\sqrt{m^2+|\bm p|^2}$ is the on-shell condition and the sum over $N$ goes from $s$ to $-s$, which are all the $2s+1$ physical states.
One then defines the time-ordered Green's function as
\begin{eqnarray}
\nonumber
\mathbb {G}_{0\,MN}^{(s)}(\bm x,\bm x', t,t')=\langle 0| {\cal T}\bar\Psi^{(s)}(\bm x,t)\Psi^{(s)}(\bm x',t')|0\rangle\,,\\
\end{eqnarray}
where we have defined $\bar \Psi^{(s)}\equiv \beta \Psi^{(s)\dagger}$ and $\beta$ is given by 
\begin{eqnarray}
\beta= \left[\begin{array}{cc} 0 & \mathbb I_{(2s+1)\times (2s+1)} \\ \mathbb I_{(2s+1)\times (2s+1)}  &0  \end{array} \right]\,.
\end{eqnarray}
Using (\ref{left and right field expansions in flat space}) and the commutation relations 
\begin{eqnarray}
\left[a_{\bm p,M},a^\dagger_{\bm p',M'} \right]_{\pm}=(2\pi)^3\delta^3(\bm p-\bm p')\delta_{MM'}\,, \quad \left[b_{\bm p,M},b^\dagger_{\bm p',M'} \right]_{\pm}=(2\pi)^3\delta^3(\bm p-\bm p')\delta_{MM'}\,,
\label{commutations of a and b in general}
\end{eqnarray}
one can easily show
\begin{eqnarray}
\mathbb {G}_{0}^{(s)}(\bm x,\bm x', t,t')=\int \frac{d^3 p}{(2\pi)^{3}} e^{i \bm p\cdot \left(\bm x-\bm x'\right)} \mathscr{G}_0(\bm p, t,t')\,,
\label{massive Dirac Greens functions}
\end{eqnarray}
where
\begin{eqnarray}
\mathscr{G}_0(\bm p, t,t')=\frac{1}{2\omega(\bm p)}\left\{ \begin{array}{cc} \left[\begin{array}{cc} \mathbb I& (m)^{-2s}{\Pi}(p^0,{\bm p})\\ (m)^{-2s}\bar{\Pi}(p^0,{\bm p})&\mathbb I\end{array}\right]e^{-i\omega(\bm p)(t-t')}\,,& t>t'\\ (-1)^{2s} \left[\begin{array}{cc} \mathbb I& (m)^{-2s}{\Pi}(p^0,-{\bm p})\\ (m)^{-2s}\bar {\Pi}(p^0,-{\bm p})&\mathbb I\end{array}\right]e^{-i\omega(\bm p)(t'-t)}
\,, & t<t' \end{array}  \right.\,.
\label{momentum massive Dirac Greens functions}
\end{eqnarray}
The polarizations  $\Pi$ and $\bar \Pi$ are the time-like non-normalized version of $\widehat \Pi$ defined in (\ref{Pi in terms of p}). They were derived in \cite{Weinberg:1964cn} and we do not repeat this derivation here. They take the form
\begin{eqnarray}
\Pi(p^0,\bm p)=(-1)^{2s}t^{a_1...a_{2s}}p_{a_1}...p_{a_{2s}}\,, \quad \bar\Pi(p^0,\bm p)=(-1)^{2s}\bar t^{a_1...a_{2s}}p_{a_1}...p_{a_{2s}}
\label{general forms of Pi}
\end{eqnarray}
where $p^\mu$ are time-like on-shell vectors, and the tensors $t^{a_1...a_{2s}}$ and $\bar t^{a_1...a_{2s}}$ are defined in Section (\ref{Spinor calculus}).

\subsection{Spinor calculus}
\label{Spinor calculus}

In this section we pause to summarize an important piece of group theory that we have been using throughout this work. In particular, we elucidate the philosophy behind the generalized Pauli matrices $t^{a_1...a_{2s}}$ and $\bar t^{a_1...a_{2s}}$ that appeared in previous sections. 

We start with the usual Pauli matrices $\bm t\equiv \bm \sigma$ and $t^0\equiv \sigma^0={\mathbb I}$, which transform as  four-vectors in the sense
\begin{eqnarray}
D^{(1/2)}\left[\Lambda\right]t^{a}D^{(1/2)\dagger}\left[\Lambda\right]=\Lambda_{b}^a t^b\,.
\end{eqnarray} 
where $\Lambda$ is a general proper Lorentz transformation and  $D^{(1/2)}\left[\Lambda\right]$ is the corresponding $2\times 2$ Lorentz transformation matrix in the $(1/2,0)$ representation. This is the familiar construction of vectors in the basis of the defining representation $(1/2,0)$ of $su(2)$. Similarly, the matrices $\bar{\bm t}\equiv -\bm \sigma$ and $\bar t^0\equiv \sigma^0={\mathbb I}$ transform as  vectors in the $(0,1/2)$ representation. In fact, one can generalize this construction to represent tensors of rank $2s$ using the $2s+1$ dimensional representation matrices \cite{Barut:1963zzb,Weinberg:1964cn,Weinberg:1964ev}.  Then, one can prove that  for irreducible representations of the Lorentz algebra there exists a set of $(2s+1)\times (2s+1)$ dimensional matrices $t^{a_1a_2...a_{2s}}$ such that: (1) $t^{a_1a_2...a_{2s}}$ is symmetric in $a_1,a_2,...,a_{2s}$, (2)  traceless in all indices, i.e. $\eta_{a_1a_2}t^{a_1a_2...a_{2s}}=0$, and that (3) $t^{a_1a_2...a_{2s}}$ transforms as 
\begin{eqnarray}
D^{(s)}[\Lambda]t^{a_1a_2...a_{2s}}_{MN}D^{(s)\dagger}[\Lambda]=\Lambda_{a_1}^{b_1}\Lambda_{a_2}^{b_2}...\Lambda_{a_2s}^{b_2s}t^{b_1b_2...b_{2s}}_{MN}\,,
\label{trans law t}
\end{eqnarray}
where $M,N=s,s-1,...,-s+1,-s$, and $D^{(s)}\left[\Lambda\right]$ is the corresponding $(2s+1)\times (2s+1)$ Lorentz transformation matrix in the $(s,0)$ representation \cite{Weinberg:1964cn,Weinberg:1964ev}.  Similarly, there exists a set of $(2s+1)\times (2s+1)$ dimensional matrices $\bar t^{a_1a_2...a_{2s}}$ that correspond to tensors in the $(0,s)$ representation. They transform according to
\begin{eqnarray}
D^{(s)\dagger}[\Lambda^{-1}]\bar t^{a_1a_2...a_{2s}}_{MN}D^{(s)}[\Lambda^{-1}]=\Lambda_{a_1}^{b_1}\Lambda_{a_2}^{b_2}...\Lambda_{a_2s}^{b_2s}\bar t^{b_1b_2...b_{2s}}_{MN}\,.
\label{trans law tbar}
\end{eqnarray}
One can show that
\begin{eqnarray}
\bar t^{a_1a_2...a_{2s}}=(\pm 1)t^{a_1a_2...a_{2s}}\,,
\end{eqnarray}
where the sign is $+1$ or $-1$ according to whether there are an even or odd number of space-like indices, respectively. Also one can show  \cite{Weinberg:1964cn,Weinberg:1964ev}
\begin{eqnarray}
t^{(a_1a_2...a_{2s}}\bar t^{b_1b_2...b_{2s})}=\eta^{(a_1b_1}...\eta^{a_{2s}b_{2s})}\,,
\end{eqnarray}
where the parentheses denote complete symmetrization over the indicated indices. One can also prove the trace identity
\begin{eqnarray}
\mbox{tr}\left[t^{a_1a_2...a_{2s}}\bar t^{b_1b_2...b_{2s}}\right]=C_1 \eta^{a_{i_1}a_{j_1}}...\eta^{b_{i_2}b_{j_2}}+C_2 \eta^{a_{i_1}b_{j_1}}...\eta^{b_{i_2}a_{j_2}}+...\,,
\label{tr contraction}
\end{eqnarray}
where $\{C_i\}$ are constants. For example, in the case $s=1$ we have $\mbox{tr}\left[\bar t^{a_1 a_2}t^{b_1b_2}\right]=-g^{
a_1 a_2}g^{b_1 b_2}+2g^{a_1 b_1}g^{a_2 b_2}+2g^{a_1b_2}g^{a_2 b_1}$.  The trace identity will be crucial to our construction as we show in Section \ref{Vertex operators}. 

Since the set of matrices  $\{t^{a_1a_2...a_{2s}}\}$ are linearly independent, we can use them to express $\widehat{\Pi_{MN}}(\hat{\bm p})=D_{M,-s}^{(s)}[R(\hat p)]D_{N,-s}^{*(s)}[R(\hat p)]$ in a covariant form:
\begin{eqnarray}
\widehat{\Pi_{MN}}(p^0,{\bm p})=\frac{(-1)^{2s}}{|2\bm p|^{2s}}t_{MN}^{a_1a_2...a_{2s}}p_{a_1}p_{a_2}...p_{a_{2s}}\,,
\end{eqnarray}
where $p$ is a light-like vector. Finally, defining $n_a(\bm p)\equiv \left(1, \frac{\bm p}{|\bm p|}\right)$ we can write $\widehat{\Pi_{MN}}(\hat{\bm p})$ that we defined in (\ref{Pi in terms of p}) as
\begin{eqnarray}
\widehat{\Pi_{MN}}(p^0,{\bm p})=\frac{(-1)^{2s}}{2^{2s}}t_{MN}^{a_1a_2...a_{2s}}n_{a_1}(\bm p)n_{a_2}(\bm p)...n_{a_{2s}}(\bm p)\,.
\label{Pi in terms of t}
\end{eqnarray}
By comparing (\ref{Pi in terms of p}) and (\ref{Pi in terms of t}) we can read off the expressions of $t_{MN}^{a_1a_2...a_{2s}}$ in terms of $\{J^i\}$. For example, for the case $s=1$ we have $t^{00}=\bar t^{00}={\cal I}\,,t^{0i}=t^{i0}=\bm J_i=-\bar t^{i0}\,,t^{ij}=\bar t^{ij}=\{\bm J_i,\bm J_j\}-\delta_{ij}$. We can use the same matrices to write the covariant expression of $\Pi(p^0,\bm p)$ given in (\ref{general forms of Pi}).

\subsection{General causal fields in curved space: the vierbein formalism}
\label{General causal fields in curved space}

The study of higher spin fields in curved background is of paramount importance since the production of ultra heavy higher spin resonances in expanding background could leave  a distinct feature in the large scale observables, e.g., energy-momentum tensor correlators.  In this section we review the {\em vierbein formalism}, which is necessary to construct general causal fields in curved background. Although our construction is general enough, the present work will limit the treatment to $s=1/2$ particles in curved background. Higher spin particles in curved background will be pursued elsewhere.  

Vierbeins, which are denoted by $V_{\mu}^a$, are tangent vectors to a set of locally inertial coordinates $\xi^a_X$   erected at a spacetime point $X$:
\begin{eqnarray}
V_{\mu}^a\equiv \left(\frac{\partial\xi^a}{\partial x^\mu}\right)_{x=X}\,.
\end{eqnarray}
The Greek letters denote the curved space coordinates, while the latin letters $a,b,..$ denote the flat space coordinates, which are raised and lowered with the flat space metric $\eta_{\mu\nu}$.  We fix $\xi^a_X$ at each spacetime point such that under a general non-inertial coordinate transformation we have
\begin{eqnarray}
V'^a_{\mu}=\frac{\partial x^\nu}{\partial x'^\mu}V^a_\nu\,.
\end{eqnarray}
$V^a_\mu$ can be thought of $4$ different covariant vectors, one for each value of $a$. Therefore, we refer any four vector or tensor at point $x$ to an inertial frame at the same point by contracting it with the vierbein. For example, the curved spacetime metric tensor $g^{\mu\nu}(x)$ can be contracted with two virebians to obtain the inertial frame metric $\eta_{ab}$ as follows 
\begin{eqnarray}
\eta^{ab}=V^{a}_{\mu}(x)V^{b}_{\nu}(x)g^{\mu\nu}(x)\,
\end{eqnarray}
or the inverse relation
\begin{eqnarray}
g_{\mu\nu}(x)=V_{\mu}^a(x) V_{\nu}^b(x) \eta_{ab}\,.
\end{eqnarray}
Using this language we classify various objects as being Lorentz  scalars, spinors, vectors, etc., or coordinate scalars, spinors, vectors, etc., or mixed combinations under both Lorentz and coordinate transformations. For example,  $V_{\mu}^a$ is a Lorentz vector and coordinate vector, $g_{\mu\nu}$ is a coordinate tensor and Lorentz scalar, while $\eta_{ab}$ is a coordinate scalar and Lorentz tensor.  
 
One uses this technology to construct physical quantities (like the general causal fields or their covariant derivatives for example) at inertial frames and then refer them to general spacetime points using veirbiens. Thus, one can show that the covariant derivative ${\cal D}_a$ is given by \cite{Weinberg:1972kfs}
\begin{eqnarray}
{\cal D}_a=V_{a}^\mu \left(\partial_\mu+\Gamma_\mu\right)\,,
\end{eqnarray}
where the spin connection $\Gamma_\mu$ is 
\begin{eqnarray}
\Gamma_\mu=\frac{1}{2}{\cal T}^{ab}V_a^\nu V_{b\nu;\mu}\,,
\end{eqnarray}
and ${\cal T}^{ab}$ are the corresponding Lorentz generators of the object being differentiated. Here, one has to be carefule when dealing with a set of nested Lorentz generators.   For example, the left-handed massless field $\Phi^{(s)}_L$ transforms as $(s,0)$, while the covariant derivative ${\cal D}_{b}$ transforms as $\left(\frac{1}{2},\frac{1}{2}\right)$. Thus,  ${\cal D}_{b}\Phi^{(s)}_L$ transforms as $\left(\frac{1}{2},\frac{1}{2}\right)\oplus (s,0)$. If we denote the generators and spin connection of $\left(\frac{1}{2},\frac{1}{2}\right)\oplus(s,0)$  representation by $\left[{\cal T}_{ab}\right]_{d M}^{cN}$ and $\left[\Gamma_\mu\right]_{dM}^{cN}$, respectively, then we can write them in terms of the corresponding $\left(\frac{1}{2},\frac{1}{2}\right)$  and $(s,0)$ quantities as   \cite{Birrell:1979jd}
\begin{eqnarray}
\nonumber
\left[{\cal T}_{ab}\right]_{dM}^{cN}&=&\delta_{M}^N\left[{\cal T}_{ab} \right]_{d}^c+\delta_{d}^c\left[{\cal T}_{ab} \right]_M^N\,,\\
\left[\Gamma_{\mu}\right]_{dM}^{cN}&=&\delta_{M}^N\left[\Gamma_{\mu} \right]_{d}^c+\delta_{d}^c\left[\Gamma_{\mu} \right]_M^N\,,
\end{eqnarray}
where the latin indices $a,b,c,d$ label the $\left(\frac{1}{2},\frac{1}{2}\right)$  representation, while the latin indices $M,N$ label the  $(s,0)$ representation. In particular we have for the $\left(\frac{1}{2},\frac{1}{2}\right)$ representation 
\begin{eqnarray}
\left[{\cal T}_{ab} \right]_{d}^c=\delta_{a}^d\eta_{bc}-\delta_{b}^c\eta_{ad}\,.
\end{eqnarray}
Using this construction, the massive Klein-Gordon equation of the generalized Dirac's field $\Psi^{(s)}$ takes the form
\begin{eqnarray}
\nonumber
(\Box-m^2)\Psi^{(s)}&=&\left[{\cal D}_a{\cal D}^a-m^2\right]\Psi^{(s)}_N\\
&=&V_{a}^\mu\left[\delta_{d}^a\delta^M_N\partial_\mu+\left[\Gamma_\mu\right]_{dN}^{aM} \right]V^d_\nu \left[\delta^P_M\partial^\nu+\left[\Gamma^\nu\right]_M^P \right]\Psi^{(s)}_P-m^2\Psi^{(s)}_N=0\,.
\end{eqnarray}
One can also conformally couple $\Psi^{(s)}$ (in the $m=0$ limit) by adding the term $\xi R \Psi^{(s)}$ to the Klein-Gordon equation, where $\xi=(s+1)/6$ \cite{Birrell:1979jd}. 

In this work we are particularly interested in de Sitter space:
\begin{eqnarray}
ds^2=\frac{1}{H^2\tau^2}\left(-d\tau^2+dx^2+dy^2+dz^2\right)\,,
\end{eqnarray}
with the metric
\begin{eqnarray}
g_{\mu\nu}=\frac{1}{H^2\tau^2}\eta_{\mu\nu}\,,
\end{eqnarray}
where $H$ is the Hubble's parameter and $\tau$ is the conformal time. de Sitter space is conformally flat, and hence, the easiest way to calculate the covariant derivative is to apply a Weyl transformation to the flat space such that
\begin{eqnarray}
g_{\mu\nu}\rightarrow \bar g_{\mu\nu}=\Omega^2 \eta_{\mu\nu}\,,
\end{eqnarray}
where $\Omega=\frac{1}{H\tau}$. Then, we have
\begin{eqnarray}
\nonumber
{\cal D}_a\rightarrow \bar{\cal D}_a&=&\bar V^{\mu}_a\left[\partial_\mu +\bar\Gamma_\mu \right]\,,\\
\Gamma_\mu \rightarrow \bar \Gamma_\mu&=&\Gamma_\mu-\Omega^{-1}{\cal T}^{ef}V^\nu_eV_{f\mu}\Omega_{,\nu}\,,
\end{eqnarray}
where $\bar V^{u}_a=\Omega^{-1}V_{a}^\mu$, $V_a^{\mu}=\delta_a^\mu$, and $\Gamma_\mu=0$. 

With the aid of Weyl transformation one can show that the Klein-Gordon equation takes the form 
\begin{eqnarray}
\left[\tau^2 \partial_a\partial^a-2\tau \frac{\partial}{\partial\tau}+2\tau{\cal T}^{0a}\partial_a+12\xi+\frac{m^2}{H^2}-s(s+1)\right]\Psi^{(s)}(x)=0\,.
\label{simplified equation of psi}
\end{eqnarray}
de Sittes space admits $SO(3)$ as a subgroup, and hence, one can write the Fourier transform of the three-dimensional part of the positive and negative frequency modes  of $\Psi^{(s)}$ as
\begin{eqnarray}
\Psi^{(s)}_{\pm}(\tau, \bm x)=\int \frac{d^3 p}{(2\pi)^3} e^{\pm i\bm p \cdot \bm x } \tilde\Psi^{(s)}_{\pm}(\tau, \bm p).
\label{Fourier of Psi}
\end{eqnarray}
Then, one substitutes (\ref{Fourier of Psi}) into (\ref{simplified equation of psi}) to find that the quantum field $\Psi^{(s)}$ can be expanded in terms of creation and annihilation operators as (see \cite{Birrell:1979jd,Castagnino:1986zw,Grensing:1977fn} for details) 
\begin{eqnarray}
\nonumber
\Phi_{L\,,M}^{(s)}&=&\int \frac{d^3 p}{(2\pi)^3}\sum_{N}\left\{\Xi^{+L}_{MN}(\bm p, \tau)a(\bm p, N)e^{i\bm p\cdot \bm x}+\Xi^{-L}_{MN}(\bm p,\tau)b^\dagger(\bm p,N)e^{-i \bm p\cdot \bm x} \right\}\,,\\
\nonumber
\Phi_{R\,,M}^{(s)}&=&\int \frac{d^3 p}{(2\pi)^3}\sum_{N}\left\{\Xi^{+R}_{MN}(\bm p,\tau)a(\bm p, N)e^{i\bm p\cdot \bm x}+\Xi^{-R}_{MN}(\bm p,\tau)b^\dagger(\bm p,N)e^{-i\bm p\cdot \bm x} \right\}\,,\\
\label{left and right field expansions}
\end{eqnarray} 
where
\begin{eqnarray}
\nonumber
&&\left[\begin{array}{c} \Xi^{+L}_{MN}\\ \Xi^{+R}_{MN}\end{array}\right]=\frac{H\tau}{2\sqrt{|\bm p|}}\left[\begin{array}{c}e^{\frac{i\pi N}{2}}\left(\frac{m}{H}\right)^{N}W_{-N,\nu}\left(2i|\bm p|\tau\right)D_{M,N}^{(s)}\left[R(\hat {\bm p})\right] \\
e^{\frac{-i\pi N}{2}}\left(\frac{m}{H}\right)^{-N}W_{N,\nu}\left(2i|\bm p|\tau\right)D_{M,N}^{(s)}\left[R(\hat {\bm p})\right] 
 \end{array}\right]\,,\\
&&\left[\begin{array}{c} \Xi^{-L}_{MN}\\ \Xi^{-R}_{MN}\end{array}\right]=\frac{H\tau}{2\sqrt{|\bm p|}}\left[\begin{array}{c}e^{\frac{-i\pi N}{2}}\left(\frac{m}{H}\right)^{N}W_{-N,\nu}\left(-2i|\bm p|\tau\right)D_{M,N}^{(s)}\left[R(\hat {\bm p})\right] \\
(-1)^se^{\frac{-i3\pi N}{2}}\left(\frac{m}{H}\right)^{-N}W_{N,\nu}\left(-2i|\bm p|\tau\right)D_{M,N}^{(s)}\left[R(\hat {\bm p})\right] 
 \end{array}\right]\,,
\label{various Xi functions}
\end{eqnarray}
$\nu^2=\frac{9}{4}-m^2H^2-12\xi+s(s+1)$, and $W_{a,b}(x)$ are the Whittaker's functions. 

In this work we will only consider left-handed massless fields in the $s=1/2$ representation. In this case the sum over $N$ in (\ref{left and right field expansions}) is restricted to a single value $N=-s$. 

Next, we define the Green's functions of the left fields as (with a similar form of the right field Green's function after replacing $\Pi$ with $\bar \Pi$)
\begin{eqnarray}
\nonumber
G_{0\,MN}^{(s)}(\bm x,\bm x', \tau,\tau')&=&\langle 0| {\cal T}\Phi^{(s)}_M(\bm x,\tau)\Phi^{(s)\dagger}_N(\bm x',\tau')|0\rangle\\
&=& \int \frac{d^3p}{(2\pi)^3}e^{i\bm p\cdot \left(\bm x-\bm x'\right)}\left\{\begin{array}{cc} \widehat{\Pi_{MN}}(p^0,{\bm p})u^*(\bm p,\tau')u(\bm p,\tau)\,, & \tau>\tau'\\ \widehat{\Pi_{MN}}(p^0,-{\bm p})v^*(-\bm p,\tau')v(-\bm p,\tau)\,, & \tau'>\tau \end{array} \right.\,, 
\label{curved space Greens functions}
\end{eqnarray} 
where 
\begin{eqnarray}
\nonumber
u(\bm p, \tau)&=&\frac{H\tau}{2\sqrt{|\bm p|}}e^{\frac{i\pi s }{2}}\left(\frac{m}{H}\right)^{s}W_{s,\nu}\left(-2i|\bm p|\tau\right)\,,\\
v(\bm p, \tau)&=&\frac{H\tau}{2\sqrt{|\bm p|}}e^{\frac{-i\pi s }{2}}\left(\frac{m}{H}\right)^{s}W_{s,\nu}\left(2i|\bm p|\tau\right)\,.
\end{eqnarray}
One can easily show that the expression (\ref{curved space Greens functions}) degenerates to (\ref{momentum space left handed particles}) in the limit $H,m\rightarrow 0$. 

This ends all the mathematical background needed in order to compute energy-momentum tensor correlators of higher spin fields in a fixed background.  In the next sections we show how this can be achieved using general principles of covariance and analyticity.  

\section{Path integral, effective action, and energy momentum tensor}
\label{Path integral, effective action, and energy momentum tensor}

\subsection{Motivation and strategy}

The production of particles from vacuum will generally be accompanied by the generation of macroscopic energy-momentum tensor since this process puts physical states on shell. However,  this process is stochastic, and hence, the average of the transverse traceless part of the energy-momentum tensor over all directions and/or momenta is expected to vanish. Nevertheless, we still can obtain non-trivial energy-momentum tensor correlators, which may carry distinct features of the particles produced. The lack of a Lagrangian formulation of the general causal fields makes the task of obtaining an energy-momentum tensor, not to mention its correlators, a non-trivial one. 

In this section we postulate a form of the energy-momentum tensor of the general causal fields. This tensor satisfies all the conditions of an energy-momentum tensor: it is symmetric, conserved, and Lorentz covariant second rank tensor. We motivate the form of this tensor by carefully constructing the corresponding one for $s=\frac{1}{2}$ field with no reference to a Lagrangian formulation. The idea is based on the observation that the knowledge of the free-field Green's function in a fixed background can be used to define an effective action of the field.  Then, by perturbing the background with appropriate operators we can obtain the energy-momentum tensor as well as its second and higher order correlators by taking a series of functional derivatives of the effective action with respect to the background metric. In flat space, we find that the set of operators can be determined exactly by requiring the existence of non-vanishing, conserved, and covariant energy-momentum tensor. The same method can in principle be used to sort out the set of allowed operators in curved background, modulo a non-trivial point regarding the non-commutativity of covariant derivatives for spins $s>\frac{1}{2}$. In both cases (flat and curved background), we find that the use of the analytical properties of the on-shell-time-ordered Greens' function, i.e., the on-shell Feynman propagator, is enough  to determine the energy-momentum tensor correlators that emerge due to production of the general causal fields from vacuum. 

We first apply our formalism to the massless case in flat background, where algebra is the least cumbersome. Then, we show that the same formalism can be generalized to the massive case and in curved background.  

\subsection{Massless fields}

Given an action $S[\Phi, \Phi^\dagger]$, where $\Phi$ is any complex field, e.g., the spin $\frac{1}{2}$ field, then one can write the path integral as (in the following we keep the spin index $s$ general for later convenience)
\begin{eqnarray}
Z[J]=\int {\cal D}[\Phi] {\cal D}[\Phi^\dagger]e^{iS[\Phi,\Phi^\dagger]+i\int d^4x\sqrt{-g(x)} J(x) \Phi(x)-i\int d^4x\sqrt{-g(x)}  J^\dagger(x) \Phi^\dagger(x)}\,,
\label{path integral}
\end{eqnarray}
in the presence of an external current $J$. Physically, the partition function $Z$ is  the vacuum persistence amplitude $\langle \mbox{Out},0|0,\mbox{In} \rangle$. The presence of an external current $J$ can cause an instability in the initial vacuum state $|0,\mbox{In}\rangle$ leading to particle production. In the absence of currents in flat space we have $|0,\mbox{In}\rangle=|0,\mbox{Out}\rangle=|0\rangle$, and hence $Z[0]=\langle 0|0\rangle=1$ and no particle production can take place. However, particle production can occur in curved spacetime since in general $|0,\mbox{In}\rangle \neq|0,\mbox{Out}\rangle$ even in the absence of external current \cite{Birrell:1982ix}.  Therefore, one can mimic the effect of curved spacetime by turning on an external current in flat space and vise versa.  We also define the effective action $W$ as
\begin{eqnarray}
Z[0]=\langle \mbox{Out},0|0, \mbox{In} \rangle\equiv e^{iW}\,.
\label{definition of W}
\end{eqnarray} 
In fact, the effective action provides us with a valuable tool to compute the energy momentum tensor and its correlators, as we will see in this section. This is particularly important in our case because of the absence of a Lagrangian formulation for the general causal fields.
To this end we set $J=0$ and examine the variation of  (\ref{path integral}) to find
\begin{eqnarray}
\delta Z[0]=i\int {\cal D}[\Phi] {\cal D}[\Phi^\dagger]\delta S e^{iS}=i \langle \mbox{Out},0|\delta S|0, \mbox{In} \rangle\,,
\end{eqnarray}
from which one obtains
\begin{eqnarray}
\langle \mbox{Out},0|T_{\mu\nu}|0, \mbox{In} \rangle=-i\frac{2}{\sqrt{-g}}\frac{\delta Z[0]}{\delta g_{\mu\nu}}\,.
\end{eqnarray}
Finally, one can use (\ref{definition of W}) to find:
\begin{eqnarray}
\nonumber
\langle T_{\mu\nu}\rangle&\equiv&\frac{\langle \mbox{Out},0|T_{\mu\nu}|0, \mbox{In} \rangle}{\langle \mbox{Out},0|0, \mbox{In} \rangle}=\frac{2}{\sqrt{-g}}\frac{\delta W}{\delta g^{\mu\nu}}\,,\\
\langle T_{\mu\nu}(x)T_{\alpha\beta}(y)\rangle&=&\frac{4}{\sqrt{-g(x)}\sqrt{-g(y)}}\frac{\delta^2 W}{\delta g^{\mu\nu}(x)\delta g^{\alpha\beta}(y)}\,.
\label{definition of energy momentum tensors using W}
\end{eqnarray}

The partition function of the free field $\Phi$ in a fixed curved background is given by
\begin{eqnarray}
Z[0]=\int {\cal D}[\Phi] {\cal D}[\Phi^\dagger]e^{i \int d^4 x\sqrt{-g(x)}\Phi^\dagger(x)\left[G^{-1}(x,x')\right]_{x'\rightarrow x}\Phi(x)}\,.
\end{eqnarray}
Since the integral over $\Phi$ is quadratic, it can be readily performed to obtain:
\begin{eqnarray}
Z[0]=\mbox{Det}\left[G^{-1}\right]^{(-1)^{2s+1}}\,.
\end{eqnarray}
Now, using (\ref{definition of W}) we find
\begin{eqnarray}
\nonumber
W&=&-i(-1)^{2s+1}\mbox{Tr}\log\left[G^{-1}\right]\\
&=&-i(-1)^{2s+1}\int d^4x\sqrt{-g(x)}\mbox{tr}\log\left[G^{-1}(x,x')\right]_{x'\rightarrow x}\,,
\label{definition of effective action}
\end{eqnarray}
where  Tr denotes the trace over spacetime and Lorentz indices, while tr denotes the trace over Lorentz indices only. Next, we consider a small perturbation in the background metric:
\begin{eqnarray}
g_{\mu\nu}=g_{0\,\mu\nu}+\delta g_{\mu\nu}\,,\quad \delta\sqrt{-g}=\frac{1}{2}\sqrt{-g_0}g^{\mu\nu}_0\delta g_{\mu\nu}\,,
\end{eqnarray}
where $g_{0\,\mu\nu}$ is the unperturbed metric. 
The Green's function responds to the perturbation of $g$ by acquiring an extra piece
\begin{eqnarray}  
G^{-1}=G^{-1}_0+\delta g_{\mu\nu}{\cal O}^{\mu\nu}\,,
\end{eqnarray}
where $G_0$ is the unperturbed Green's function and ${\cal O}^{\mu\nu}$ is the vertex operator that will be determined in the next section. Using the cyclic property of the trace we have \newline $\mbox{tr}\log\left[ G^{-1}\right]=\mbox{tr}\log \left[G_0^{-1}\left[I+\delta g_{\mu\nu}{\cal O}^{\mu\nu}G_0\right]\right]$. Thus, we can write the effective action as
\begin{eqnarray}
W=-i(-1)^{2s+1}\int d^4 x\sqrt{-(g_0+\delta g)}\left\{\mbox{tr}\log\left[G_0^{-1}\right]+\mbox{tr}\log\left[I+\delta g_{\mu\nu}O^{\mu\nu} G_0 \right]\right\}\,.
\end{eqnarray}  
Expanding to second order in $\delta g_{\mu\nu}$ we obtain
\begin{eqnarray}
\nonumber
W&=&-i(-1)^{2s+1}\left\{ \int d^4x \sqrt{-g_0(x)}\mbox{tr}\log \left[G_0^{-1}(x,x') \right]_{x'\rightarrow x} \right.\\
\nonumber
&&\left. +\int d^4 x \sqrt{-g_0(x)} \delta g_{\mu\nu}(x)\mbox{tr}\left[O^{\mu\nu}(x)G_0(x,x')\right]_{x'\rightarrow x}\right.\\
\nonumber
&&\left. + \frac{1}{2}\int d^4 x \sqrt{-g_0(x)} g_0^{\mu\nu} \delta g_{\mu\nu}(x)\left[ \mbox{tr}\log\left[ G_0^{-1}\right]_{x'\rightarrow x}+\mbox{tr}\left[\delta g_{\alpha\beta}(x){\cal O}^{\alpha\beta}(x)G_0(x,x')\right]_{x'\rightarrow x} \right]\right.\\
\nonumber
&&\left.-\frac{1}{2}\int d^4x d^4 y\sqrt{-g_0(x)}\sqrt{-g_0(y)}\delta g_{\mu\nu}(x)\delta g_{\alpha\beta}(y)\mbox{tr}\left[{\cal O}^{\mu\nu}(x) G_0(x,y){\cal O}^{\alpha\beta}(y)G_0(y,x)\right]                \right\}\,.\\
\end{eqnarray}
Now, we use (\ref{definition of energy momentum tensors using W}) to find the expectation value of the energy-momentum tensor of the field $\Phi$
\begin{eqnarray}
\langle T^{\mu\nu} \rangle=-2i(-1)^{2s+1}\left\{\mbox{tr}\left[{\cal O}^{\mu\nu}(x)G_0(x,x')\right]_{x'\rightarrow x}-\frac{1}{2}g_0^{\mu\nu}\mbox{tr}\log\left[G_0(x,x')\right]_{x'\rightarrow x} \right\}\,.
\label{definition of energy momentum using the effective action}
\end{eqnarray}
The effective action (\ref{definition of effective action}) is coordinate and Lorentz scalar (in the veirbien sense), and hence, it is evident that the energy-momentum tensor is conserved; see \cite{Weinberg:1972kfs} for the proof. 
Using the same technology, we can also calculate the connected part of the energy-momentum tensor correlator: 
\begin{eqnarray}
\langle T^{\mu\nu}(x)T^{\alpha\beta}(y) \rangle_c=2i(-1)^{2s+1}\mbox{tr}\left[{\cal O}^{\mu\nu}(x)G_0(x,y){\cal O}^{\alpha\beta}(y)G_0(y,x) \right]\,.
\label{energy momentum correlator}
\end{eqnarray}

In the following we will  be interested in the momentum-space energy-momentum tensor correlator.  To this end we assume that the spacetime admits $SO(3)$ as an isometry subgroup.  Then, we can write the Green's function in Fourier space as
\begin{eqnarray}
G_0(x,y)=\int \frac{d^3 p}{\left(2\pi\right)^3}e^{i\bm p\cdot (\bm x-\bm y)}{\cal G}_0(\bm p,\tau_x,\tau_y)\,,
\end{eqnarray} 
where $\tau$ refers to either the cosmic or conformal time.  
We also use the fact that ${\cal O}^{\mu\nu}$ is a differential operator, as we will show in the next section, and take the Fourier transform of $\langle T^{\mu\nu}(x)T^{\alpha\beta}(y) \rangle_c$ to find
\begin{eqnarray}
\nonumber
\langle T^{\mu\nu}(\bm k, \tau_x)T^{\alpha\beta}(\bm k', \tau_y) \rangle_c&=&2i(-1)^{2s+1}\delta^{3}\left(\bm k+\bm k'\right)\int d^3 p~\mbox{tr}\left[{\cal O}^{\mu\nu}(\bm p,\tau_x){\cal G}_{0}(\bm p, \tau_x,\tau_y)\right.\\
&&\left. {\cal O}^{\alpha\beta}(\bm p-\bm k,t_y){\cal G}_0(\bm p-\bm k,\tau_y,\tau_x) \right]\,.
\label{definition of energy-momentum correlator}
\end{eqnarray}
Equation (\ref{definition of energy-momentum correlator}) is one of our main results in this work.

The derivation of (\ref{definition of energy momentum using the effective action}) and (\ref{energy momentum correlator}) was carried out for $s=\frac{1}{2}$ field. Since the second rank tensor in (\ref{definition of energy momentum using the effective action}) satisfies all the requirements of an energy-momentum tensor, we postulate that this definition is also valid for any $s>\frac{1}{2}$.

{\it Postulate: The energy-momentum tensor and its two-point function of the general causal fields are given by expressions (\ref{definition of energy momentum using the effective action}) and (\ref{energy momentum correlator}) (or equivalently  (\ref{definition of energy-momentum correlator}) for spaces with $SO(3)$ isometry), respectively}. Generalizing (\ref{definition of energy-momentum correlator}) to multi-point functions is a straightforward task. 

 We will show that these expressions along with the analytic properties of ${\cal G}_0(\bm p-\bm k,\tau_y,\tau_x)$ is enough to determine the large scale correlators of the energy-momentum tensor due to the production of particles from vacuum.

\subsection{Massive fields}

Turning on a mass for the fields does not change any of the above steps we followed to derive the energy-momentum tensor and its correlators. Repeating the above procedure in the massive case, where we use the Dirac representation $(s,0)\oplus(0,s)$,  we find 
\begin{eqnarray}
\langle T^{\mu\nu} \rangle=-2i(-1)^{2s+1}\left\{\mathbb{TR}\left[{\cal O}^{\mu\nu}(x)\mathbb G_0(x,x')\right]_{x'\rightarrow x}-\frac{1}{2}g_0^{\mu\nu}\mathbb{TR}\log\left[\mathbb G_0(x,x')\right]_{x'\rightarrow x} \right\}\,,
\label{definition of energy momentum in the massive case using the effective action}
\end{eqnarray}
and 
\begin{eqnarray}
\nonumber
\langle T^{\mu\nu}(\bm k, \tau_x)T^{\alpha\beta}(\bm k', \tau_y) \rangle_c&=&2i(-1)^{2s+1}\delta^{3}\left(\bm k+\bm k'\right)\int d^3 p~\mathbb{TR}\left[{\cal O}^{\mu\nu}(\bm p,\tau_x){\mathscr G}_{0}(\bm p, \tau_x,\tau_y)\right.\\
&&\left. {\cal O}^{\alpha\beta}(\bm p-\bm k,\tau_y){\mathscr G}_0(\bm p-\bm k,\tau_y,\tau_x) \right]\,,
\label{definition of energy-momentum correlator for the massive case}
\end{eqnarray}
where $\mathbb G_0$ and ${\mathscr G}_0$ in flat space are given by  (\ref{massive Dirac Greens functions}) and (\ref{momentum massive Dirac Greens functions}), respectively, and $\mathbb{TR}$ is the trace over Dirac indices in the $(s,0)\oplus(0,s)$ representation. 

\section{Vertex operators}
\label{Vertex operators}

In the previous section we showed how one can construct the energy-momentum tensor, and its correlators, of a free general causal field in a given background. However, this construction forced us to introduce vertex operators ${\cal O}^{\mu\nu}$ that are yet to be determined. This section is devoted to elucidate the method we use to construct the vertex operators, which hinges on general considerations of Lorentz covariance. 
The general idea is that ${\cal O}^{\mu\nu}$ is used to write  the energy-momentum tensor via (\ref{definition of energy momentum using the effective action}), and hence,  we can use the symmetries of this tensor to construct ${\cal O}^{\mu\nu}$. The construction of the vertex operators in flat space follows immediately from symmetry considerations of $T^{\mu\nu}$. The construction, however, is more involved in curved background.

\subsection{Flat space vertices: massless fields}

The energy-momentum tensor has to obey the following obvious criteria: 
(1) it transforms covariantly under Lorentz transformations, 
(2) it is conserved, and
(3) it is symmetric in its two indices. Therefore, the momentum-space expression of the energy-momentum tensor of any on-shell massless particle in flat space has to take the form
\begin{eqnarray}
\langle T^{\mu\nu}(p^2)\rangle =f_1(p^2) p^\mu p^\nu-f_2(p^2)g^{\mu\nu}\,.
\label{simple energy-momentum tensor}
\end{eqnarray}
In this section we use both the Greek $\mu,\nu$ and Latin $a,b$ letters to denote flat space coordinates. Now, since we are dealing with on-shell massless particles we have $p^2=0$, and hence, $f_1(p^2)$ and $f_{2}(p^2)$ are at most constants: $f_1(p^2)=f_1, f_2(p^2)=f_2$. However, conservation of $T^{\mu\nu}$, i.e., $p_\mu T^{\mu\nu}=0$, demands the vanishing of $f_2$.  Therefore, the energy-momentum tensor takes the simple form
\begin{eqnarray}
\langle T^{\mu\nu}(p^2)\rangle=f p^\mu p^\nu\,
\label{simple energy-momentum tensor 2}
\end{eqnarray}
or its inverse Fourier transform
\begin{eqnarray}
\langle T^{\mu\nu}(x)\rangle =f\int \frac{d^3 p}{(2\pi)^3 2 |\bm p|} e^{i\bm p \cdot \bm x-i|\bm p|t} p^{\mu}p^{\nu}
\label{simple energy-momentum tensor 3}
\end{eqnarray}
will guide us in our search for the vertex operators in flat space. 

The energy-momentum tensor defined via (\ref{definition of energy momentum using the effective action}) should coincide with (\ref{simple energy-momentum tensor 3}). Therefore, we first demand that $\mbox{tr}\log G_0=0$, which can be achieved through appropriate renormalization. Now, the trace of ${\cal O}^{\mu\nu}$ operating on $G_0$ should reproduce $\sim p^\mu p^\nu$ after applying Fourier transform. Recalling that $G_0\sim t^{a_1a_2...a_{2s}}$ and the fact $\eta_{a1a2}t^{a_1a_2...a_{2s}}=0$, etc. (see Section \ref{Spinor calculus}) forces the vertex to be an object that carries spinoral indices such that its contraction with $t^{a_1a_2...a_{2s}}$ gives a non-zero result. Also recalling the transformation laws of $t^{a_1a_2...a_{2s}}$ and $\bar t^{a_1a_2...a_{2s}}$ , see (\ref{trans law t}) and (\ref{trans law tbar}),  along with the trace identity (\ref{tr contraction}) leaves us with the only choice that the vertex ${\cal O}^{\mu\nu}$ has to be proportional to the tensor $\bar t^{\mu a_2...a_{2s}}$ if we want the trace of ${\cal O}^{\mu\nu} G_0$ to be covariant under the Lorentz transformation. This completely fixes the form of the vertex operators modulo a $p$-dependent coefficient as we show below.  In order to elucidate the construction of ${\cal O}^{\mu\nu}$  we work out $s=\frac{1}{2}$ and $s=1$ as two examples and then  give the general form of ${\cal O}_{\mu\nu}$.

First we consider $s=\frac{1}{2}$. The vertex operator takes the form\footnote{One might also want to add the term $\bar t^{a}\partial_a \eta^{\mu\nu}$ to ${\cal O}^{\mu\nu}$. However, upon taking the trace $\mbox{tr}O^{\mu\nu}G_0$ we find that this term gives $p^2=0$. The same behavior occurs for higher $s$.}
\begin{eqnarray}
{\cal O}^{\left(\frac{1}{2}\right)~\mu\nu}=-i\frac{\alpha_{\frac{1}{2}}}{2}\left(\bar t^\mu \partial^\nu+\bar t^\nu \partial^\mu\right)\,.
\end{eqnarray}
Then using  the identity $\mbox{tr}\left[\bar t^{\mu} t^\nu\right]=-g^{\mu\nu}$, we find from (\ref{definition of energy momentum using the effective action})
\begin{eqnarray}
\langle T^{\mu\nu} (x) \rangle=2i\alpha_{\frac{1}{2}}\int \frac{d^3 p}{\left(2\pi\right)^3}\frac{e^{i\bm p\cdot\bm x-ix^0|\bm p|}}{2|\bm p|}p^\mu p^\nu\,,
\end{eqnarray}
which takes the general form of the energy-momentum tensor for a massless field (\ref{simple energy-momentum tensor 3}).

The vertex for $s=1$ is
\begin{eqnarray}
{\cal O}^{(1)~\mu\nu}=-\frac{\alpha_1}{2C_1(\Box)}\left[ \bar t^{\mu\rho}\partial_\rho\partial^\nu+\bar t^{\nu\rho}\partial_\rho\partial^\mu\right]
\end{eqnarray}
Using the $(1,0)$ Green's function along with the identity $\mbox{tr}\left[\bar t^{\mu\nu}t^{\alpha\beta}\right]=-g^{
\mu\nu}g^{\alpha\beta}+2g^{\alpha\mu}g^{\nu \beta}+2g^{\mu\beta}g^{\nu \alpha}$ we find
\begin{eqnarray}
\langle T^{\mu\nu}(x)\rangle =-6i\alpha_1 \int \frac{d^3p}{(2\pi)^3}\frac{ p^2 p^\mu p^\nu}{2|\bm p|C_1(\bm p^2)}\,.
\end{eqnarray}
This expression is zero when considered on-shell. Hence, we set $C_1(p^2)=-p^2$ or $C_1(\Box)=\Box$ in order for the energy-momentum tensor to take its canonical form. 

One can easily work out the general case to find 
\begin{eqnarray}
C_s(p^2)=\left(-p^2\right)^{2s-1}\,,\quad C_s(\Box)=\Box^{2s-1}\,.
\end{eqnarray}
Therefore, the vertex takes the general form 
\begin{eqnarray}
\nonumber
{\cal O}^{(s)~\mu\nu}(x)=\frac{(-i)^{2s} \alpha_s}{2C_s(\Box)}\left[\eta_{a_1}^\mu \eta_c^\nu \bar t^{a_1 a_2...a_{2s}}\partial_{a_2}\partial_{a_3}...\partial_{a_{2s}}\partial^c+\eta_{a_1}^\nu \eta_c^\mu \bar t^{a_1 a_2...a_{2s}}\partial_{a_2}\partial_{a_3}...\partial_{a_{2s}}\partial^c \right]\,,\\
\end{eqnarray}
or in momentum space
\begin{eqnarray}
\nonumber
{\cal O}^{(s)~\mu\nu}(p)=\frac{\alpha_s}{2\left(-p^2\right)^{2s-1}}\left[\eta_{a_1}^\mu \eta_c^\nu \bar t^{a_1 a_2...a_{2s}}p_{a_2}p_{a_3}...p_{a_{2s}}p^c+\eta_{a_1}^\nu \eta_c^\mu \bar t^{a_1 a_2...a_{2s}}p_{a_2}p_{a_3}...p_{a_{2s}}p^c \right]\,,\\
\label{vertex in momentum space}
\end{eqnarray}
where $\alpha_s$ is a numerical coefficient that depends on the physics details. The appearance of the non-locality $C_s(\Box)$ for any $s>\frac{1}{2}$ is signaling a singular behavior of the massless particles, which is going to hunt us back in calculating the energy-momentum tensor correlators.

\subsection{Flat space vertices: massive fields}

We follow the same procedure to construct the vertex operators in the massive case. In particular, the most general on-shell energy-momentum tensor takes the form
\begin{eqnarray}
\langle T_{\mu\nu}(p^2)\rangle =f_1p^\mu p^\nu+f_2 \eta^{\mu\nu}\,,
\end{eqnarray}
and the conservation law $p_\mu T^{\mu\nu}=0$ gives us a non-trivial relation between $f_1$ and $f_2$. 

As we discussed before, we make use of the Dirac's representation to deal with the massive case. There are $5$ vertex operators that can  contribute to ${\cal O}^{\mu\nu}$. They can be grouped into diagonal and off diagonal operators:\footnote{We restrict our treatment to  parity conserving operators. At the end of Section \ref{Structure of the energy-momentum correlator} we comment on the fact that turning on parity violating operators and/or the over production of left over right fields (or vice versa) will lead to helical structure in the energy-momentum tensor correlators.}
\begin{eqnarray}
\nonumber
\mbox{diagonal}~&&\left\{\begin{array}{l}{\cal V}_1^{\mu\nu}=\eta^{\mu\nu}m^2\,,\\ 
{\cal V}_2^{\mu\nu}=p^\mu p^\nu\,,
\end{array} \right.\\
\mbox{off diagonal}~&&\left\{\begin{array}{l}{\cal V}_3^{\mu\nu}=\frac{(-1)^{2s}}{m^{2s-2}} p^{(\mu}\gamma^{\nu)a_2...a_{2s}}p_{a_2}...p_{a_{2s}}\,,\\
{\cal V}_4^{\mu\nu}=\frac{1}{m^{2s-2}} \eta^{\mu\nu}\gamma^{a_1a_2...a_{2s}}p_{a_1}p_{a_2}...p_{a_{2s}}\,,\\
{\cal V}_5^{\mu\nu}=  (-1)^{2s}\frac{p^\mu p^{\nu}}{m^{2s}}\gamma^{a_1...a_{2s}}p_{a_1}...p_{a_{2s}}\,.
\end{array} \right.
\end{eqnarray}
Among these operators, ${\cal V}_2$  and ${\cal V}_5$ are power counting suppressed  and we neglect them in our treatment. In addition, one can trade ${\cal V}_4$ for ${\cal V}_1$ using the generalized Dirac's equation (\ref{generalized dirac equation}).  Thus, the most general form of ${\cal O}^{\mu\nu}_{{\cal M}}$, where the subscript $_{{\cal M}}$ is used to indicate that these vertices correspond to massive particles, that yields a symmetric and conserved energy momentum tensor takes the form 
\begin{eqnarray}
\nonumber
{\cal O}^{(s)~\mu\nu}_{{\cal M}}(p)&=&\frac{\alpha_s}{({\cal S})^{2s-1}}\left[(-1)^{2s} p^{(\mu} \gamma^{\nu) a_2...a_{2s}}p_{a_2}...p_{a_{2s}}+{\cal C}_s m^{2s} \eta^{\mu\nu}{\mathbb I}_{(4s+2)\times (4s+2)}\right]\,,\\
\label{O for massive fields}
\end{eqnarray}
where ${\cal C}_s$ is a constant that we will determine momentarily, and ${\cal S}$ is some scale that has the dimension of mass square. In what follows we take ${\cal S}=m^2$ for simplicity. Using (\ref{definition of energy momentum in the massive case using the effective action}) we find
\begin{eqnarray}
\langle T^{\mu\nu}(x)\rangle=i\frac{\alpha_s}{(m^2)^{2s-1}}\int \frac{d^3 p}{(2\pi)^32\omega(\bm p)}e^{i\bm p \cdot \bm x-i\omega(\bm p)t}\mathbb{TR} \left[\begin{array}{cc} {\cal K}_{11} & {\cal K}_{12}\\ {\cal K}_{21} & {\cal K}_{22} \end{array} \right]\,,
\end{eqnarray}
where
\begin{eqnarray}
\nonumber
 {\cal K}_{11}&=& t^{c_1...c_{2s}}p_{c_1}...p_{c_{2s}}p^{(\mu}\bar t^{\nu) d_2...d_{2s}}p_{d_2}...p_{d_{2s}} +m^{4s}{\cal C}_s \eta^{\mu\nu}\mathbb I_{(2s+1)\times (2s+1)}\,,\\
{\cal K}_{12}&=& (-1)^{2s} {\cal C}_s m^{2s} t^{c_1...c_{2s}}p_{c_1}...p_{c_{2s}} \eta^{\mu\nu}+ (-1)^{2s}m^{2s} p^{(\mu}\bar t^{\nu) d_2...d_{2s}}p_{d_2}...p_{d_{2s}}\,.
\end{eqnarray}
The expressions of   ${\cal K}_{22}$ and ${\cal K}_{21}$ are obtained from  ${\cal K}_{11}$ and ${\cal K}_{12}$, receptively, by replacing $t$ by $\bar t$ and exchanging the positions of $t$ and $\bar t$. Then,  taking the trace we find
\begin{eqnarray}
\nonumber
\langle T^{\mu\nu}(p)\rangle=2i\frac{\alpha_s}{(m^2)^{2s-1}}\left\{\mbox{tr}\left[ t^{c_1...c_{2s}}p_{c_1}...p_{c_{2s}}p^{(\mu}\bar t^{\nu) d_2...d_{2s}}p_{d_2}...p_{d_{2s}}\right]   +(2s+1)m^{4s}{\cal C}_s \eta^{\mu\nu}\right\}\,.\\
\end{eqnarray}
Finally the value of ${\cal C}_s$ is determined by requiring the conservation of the energy-momentum tensor: $p_\mu T^{\mu\nu}=0$. For example ${\cal C}_{1/2}=-1/2$ and ${\cal C}_1=-1$. In fact, we will be interested only in the transverse traceless part of $T^{\mu\nu}$, and therefore, the term $m^{2s}\eta^{\mu\nu}{\mathbb I}_{(4s+2)\times (4s+2)}$  in ${\cal O}^{\mu\nu}$ can be safely neglected in our treatment.

\subsection{Curved space vertices}

Motivated by the vertex structure in flat space, we write the vertex of the massless fields in curved space as:
\begin{eqnarray}
{\cal O}^{(s)~\mu\nu}=\alpha_s\frac{(-i)^{2s}}{2C_s(\Box)}\left[V_{a_1}^\mu V_{c}^\nu+V_{a_1}^\nu V_c^\mu \right] \bar t^{a_1 a_2....a_{2s}}\llbracket D_{a_2}D_{a_3}...D_{a_{2s}}D^c\rrbracket\,,
\label{curved space vertex}
\end{eqnarray}   
where the brackets $\llbracket\,\,\rrbracket$ indicate total symmetrization of the covariant derivatives:
\begin{eqnarray}
\llbracket(D_{a_2}D_{a_3}...D_{a_{2s}}D^c\rrbracket=\sum_{\scriptsize\mbox{permutations}} D_{a_{p_1}}...D^c...D_{a_{p_{2s}}}\,.
\end{eqnarray}
This symmetrization  is dedicated by the symmetric nature of $\bar t^{a_1a_2...a_{2s}}$. The hope is that this symmetric definition will yield physical answers. Notice that unlike the flat background, the nonlocal operators $C_s(\Box)$ might not lead to singular behavior.    The treatment of massive fields in curved space follows the same lines and we refrain from discussing them further here. The investigation of all these points will be left for a future work.  In Section \ref{Energy-momentum correlators in curved space half case} we use the curved space vertex of spin $1/2$ to calculate the energy-momentum tensor correlator in FRW background as a check on our formalism.

\section{Structure of the energy-momentum tensor correlator}
\label{Structure of the energy-momentum correlator}

\begin{figure}[t] 
   \centering
   \includegraphics[width=4in]{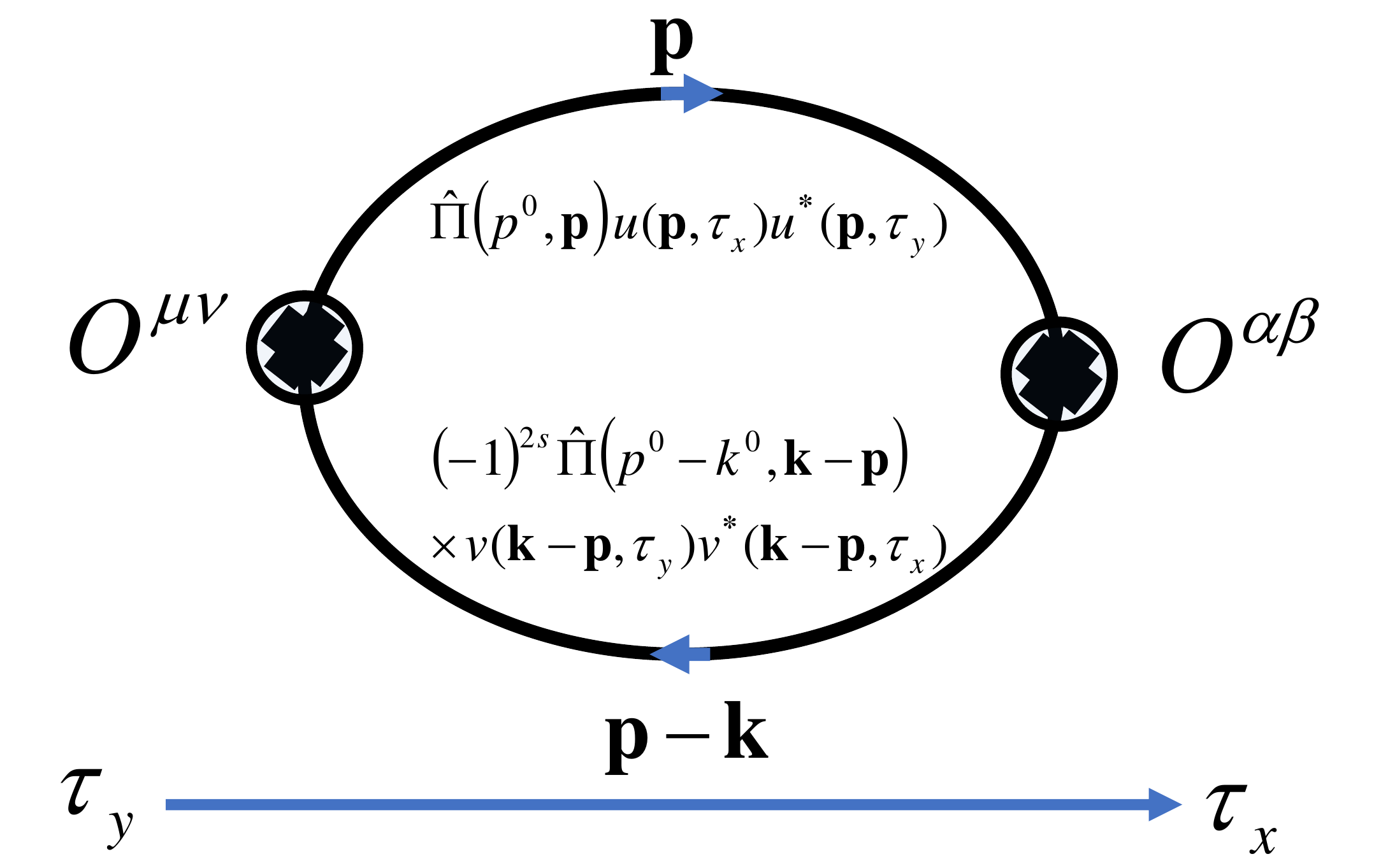} 
   \caption{The structure of the two-point function of the energy-momentum tensor. The time flows from left, $\tau_y$, to right, $\tau_x$.}
	\label{two point function of EM tensor}
\end{figure}

In the previous section we constructed the on-shell vertex operators required to compute the energy-momentum tensor and its correlators. 
Now, we are in a position to proceed to these computations. The analytic structure of the correlator is dictated by the time-ordered Green's function (\ref{time ordered Greens function}), i.e., the Feynman propagator\footnote{Once again, we stress that all our computations are done on-shell as we are interested in the on-shell contribution of the particles produced from vacuum to the energy-momentum tensor.}.  In particular,  one has to exercise caution in determining which part of the Green's function to pick, which depends on whether $\tau_x>\tau_y$ or vise versa, as we go from left to right in  (\ref{definition of energy-momentum correlator}) and (\ref{definition of energy-momentum correlator for the massive case}). Assuming that $\tau_x>\tau_y$ we pick a factor of $(-1)^{2s}$ from the second Green's function which cancels out with the same factor coming from the effective action. This intricate process is depicted in Figure \ref{two point function of EM tensor}.  Finally, the energy-momentum tensor correlator for the left-handed massless fields reads
\begin{eqnarray}
\nonumber
\langle T^{\mu\nu}(\bm k, \tau_x)T^{\alpha\beta}(\bm k', \tau_y) \rangle_c&=&-2i\delta^{3}\left(\bm k+\bm k'\right)\int \frac{d^3 p}{4|\bm p||\bm p-\bm k|}\\
\nonumber
&&\mbox{tr}\left[{\cal O}^{(s)~\mu\nu}(\bm p,\tau_x)\widehat{\Pi}(p^0, {\bm p})u(\bm p,\tau_x)u^*(\bm p, \tau_y)\right.\\
\nonumber
&&\left. {\cal O}^{(s)~\alpha\beta}(\bm p-\bm k,\tau_y)\widehat{\Pi}(p^0-k^0,\bm k-\bm p)v^*(\bm k-\bm p,\tau_x)v(\bm k-\bm p, \tau_y) \right]\,,\\
\label{the main EM correlator}
\end{eqnarray}
where $\widehat{\Pi}(p^0-k^0,\bm k-\bm p)$ should be understood as the polarization tensor obtained by a rotation by some angle $\theta$, about a reference axis, to the light-like vector $\bm p -\bm k$ (as usual, we take the reference axis to be the $z$-axis). The correlator of the  massless right-handed fields can be obtained by replacing $\Pi$ with $\bar \Pi$ and ${\cal O}^{(s)~\mu\nu}$ with $\bar {\cal O}^{(s)~\mu\nu}$ (this demands replacing $\bar t$ with $t$ in ${\cal O}^{(s)~\mu\nu}$). A similar procedure leads to the correlator of massive fields:
\begin{eqnarray}
\nonumber
\langle T^{\mu\nu}(\bm k, \tau_x)T^{\alpha\beta}(\bm k', \tau_y) \rangle_c&=&-2i\delta^3(\bm k+\bm k')\int \frac{d^3 p}{4\omega(\bm p)\omega(\bm p-\bm k)}\\
\nonumber
&& \mathbb{TR}\left\{{\cal O}^{(s)~\mu\nu}_{{\cal M}}(\bm p,\tau_x)\left[\begin{array}{cc} m^{2s} \mathbb I & \Pi(p^0, \bm p) \\ \bar\Pi(p^0, \bm p)& m^{2s} \mathbb I  \end{array}\right]{\cal U}(\bm p,\tau_x){\cal U}^*(\bm p, \tau_y)\right.\\
\nonumber
&&\times \left.{\cal O}^{(s)~\mu\nu}_{{\cal M}}(\bm p-\bm k,\tau_y)\left[\begin{array}{cc} m^{2s} \mathbb I & \Pi((k-p)^0, \bm k-\bm p) \\ \bar\Pi((k-p)^0, \bm k-\bm p)& m^{2s} \mathbb I  \end{array}\right]\right.\\
&&\times\left.{\cal V}^*(\bm k-\bm p,\tau_x){\cal V}(\bm k-\bm p, \tau_y)\right\}\,,
\label{the main massive EM correlator}
\end{eqnarray}
where we have used a condensed notation such that ${\cal U}$ and ${\cal V}$ can denote any of the $\Xi$ functions defined in (\ref{various Xi functions}). The explicit form of  ${\cal U}$ and ${\cal V}$ will not concern us in the present work since we limit our examination to the flat space correlators (massive and massless),  in addition to $s=\frac{1}{2}$ massless case in FRW background.  Both correlators (\ref{the main EM correlator}) and (\ref{the main massive EM correlator}) are valid in a general background, keeping in mind that the construction of ${\cal O}^{(s)~\mu\nu}$ in curved space is yet to be resolved for $s>\frac{1}{2}$. It is remarkable that the structure of (\ref{the main EM correlator}) and  (\ref{the main massive EM correlator}) is solely dependent on the principles of covariance and analyticity, bearing no reference to a specific particle physics model.  

Now, we can project the energy-momentum tensor correlator along the helicity-$2$ eigenbasis defined as \cite{Weinberg:2008zzc}
\begin{eqnarray}
\epsilon_{\mu\nu}^\lambda(\bm k)=\epsilon_\mu^\lambda(\bm k)\epsilon_\nu^\lambda(\bm k)\,,
\end{eqnarray}
where $\lambda=\pm$, $\epsilon^{\pm}_0(\bm k)=0$, and 
\begin{eqnarray}
\bm \epsilon^{\pm }_\mu(\bm k)=\frac{\hat \theta \pm i \hat \phi}{\sqrt 2}\,,\quad \hat \theta=\left(\cos\theta \cos\phi, \cos\theta \sin\phi, -\sin\theta\right)\,, \quad \hat \phi=\left(-\sin \phi, \cos\phi,0\right)\,.
\end{eqnarray}
This projection has two-fold importance. First, it enables us to find out whether the energy-momentum tensor correlators  are parity violating. This can happen if the  particle production process prefers one field type (left or right) over the other. Second, it can provide a succinct method to check the unitarity of the theory as we explain below. To this end, we define the amplitude ${\cal M}^{\lambda\,,\lambda'}$ as
\begin{eqnarray}
{\cal M}^{\lambda\,,\lambda'}(\tau_x,\tau_y)\equiv i \langle T^\lambda(\bm k, \tau_x) T^{\lambda'}(-\bm k,\tau_y) \rangle_c\,,
\label{projection correlator}
\end{eqnarray}
where 
\begin{eqnarray}
T^\lambda(\bm k,\tau_x)= \epsilon_{\mu\nu}^{-\lambda}(\bm k)T^{\mu\nu}(\bm k, \tau_x)\,,
\end{eqnarray}
which measures the amount of correlation projected along the two polarization directions.

By definition the quantities $T^{+}(\bm k,\tau_x) T^{+}(-\bm k,\tau_x)$ and $T^{-}(\bm k,\tau_x) T^{-}(-\bm k,\tau_x)$ (computed at the same time point) are positive definite, which can be easily shown using the reality of $T_{\mu\nu}(\bm x,\tau_x)$ and properties of $\epsilon_\mu(\bm k)$. Therefore, a unitary theory has to yield  positive definite correlators $\langle T^{+}(\bm k,\tau_x) T^{+}(-\bm k,\tau_x)\rangle_c$ and $\langle T_{-}(\bm k,\tau_x) T_{-}(-\bm k,\tau_x)\rangle_c$, which can readily be checked from (\ref{the main EM correlator}) or  (\ref{the main massive EM correlator}) along with (\ref{projection correlator}).

\subsection{Flat space energy-momentum tensor correlators: massless fields}

The calculations of the energy-momentum tensor correlator in flat space proceed by starting from the Green's function in flat space, keeping in mind that we are interested in the contribution to the correlator from on-shell left handed particles $(s,0)$. The on-shell condition means that we have to replace every $p^0$ by $|\bm p|$  since we are dealing with massless particles. This is particularly important when we operate with ${\cal O}^{\mu\nu}$ on the Green's functions in (\ref{definition of energy-momentum correlator}). The flat space Green's function takes the form
\begin{eqnarray}
G_0(x,x')=\int \frac{d^3 p}{(2\pi)^{3}}\left(2|\bm p|\right)^{2s-1}e^{i\bm p\cdot (\bm x-\bm x')-i|\bm p||t-t'|}{\cal G}(\bm p,t,t')\,,
\end{eqnarray}
where
\begin{eqnarray}
{\cal G}(\bm p,t,t')=\left\{\begin{array}{c} \widehat\Pi(\hat{\bm p})\,,\quad t>t'\\ (-1)^{2s} \widehat\Pi(-\hat{\bm p})\,,\quad t<t' \end{array}\right.\,.
\end{eqnarray}
Operating with ${\cal O}^{\mu\nu}$ we find
\begin{eqnarray}
\nonumber
{\cal O}^{\mu\nu}(x)G_0(x,x')=(-1)^{2s-1}\alpha_s&&\int \frac{d^3 p}{(2\pi)^{3}}\frac{2^{2s-1}|\bm p|}{\left[n^2(\bm p)\right]^{2s-1}}n^{(\nu}(\bm p) \bar t^{\mu)a_2...a_{2s}}n_{a_2}(\bm p)...n_{a_{2s}}(\bm p) \\
&&\times e^{i\bm p\cdot (\bm x-\bm x')-i|\bm p||t-t'|}{\cal G}(\bm p,t,t')\,,
\end{eqnarray}
where $n(\bm p)\equiv \left(1, \frac{\bm p}{|\bm p|}\right)$ is a null vector.

Collecting everything and substituting into (\ref{the main EM correlator}) we obtain
\begin{eqnarray}
\nonumber
\langle T^{\mu\nu}(\bm k, t_x)T^{\alpha\beta}(\bm k', t_y) \rangle_c&=&\frac{-i}{2}\alpha_s^2 \left(\bm k+\bm k'\right)\int d^3 p~ e^{i\vartheta (\bm p,\bm k,t_x,t_y)}\frac{ |\bm p||\bm p-\bm k|}{\left[n^2(\bm p)\right]^{2s-1}\left[n^2(\bm p-\bm k)\right]^{2s-1}}  \\
\nonumber
&&\times \mbox{tr}\left\{ n_{a_2}(\bm p)...n_{a_{2s}}(\bm p) n^{(\mu}(\bm p) \bar t^{\nu)a_2...a_{2s}} n_{c_1}(\bm p)...n_{c_{2s}}(\bm p) t^{c_1...c_{2s}}   \right.\\
\nonumber
&&\left.\times n_{b_2}(\bm p-\bm k)...n_{b_{2s}}(\bm p-\bm k) n^{(\alpha}(\bm p-\bm k) \bar t^{\beta)b_2...b_{2s}}\right. \\
&&\left.\times n_{d_1}(\bm k-\bm p)...n_{d_{2s}}(\bm k-\bm p) t^{d_1...d_{2s}}\right\}\,,
\label{general final form of the EM correlator}
\end{eqnarray} 
where the phase $\vartheta$ is given by
\begin{eqnarray}
\vartheta (\bm p, \bm k, t_x, t_y)=\left(|\bm p|+|\bm p-\bm k|\right)(t_y-t_x)\,.
\end{eqnarray}
For $s>\frac{1}{2}$ the correlator has the highly singular piece $\left[n^2(\bm p)\right]^{2s-1}\left[n^2(\bm p-\bm k)\right]^{2s-1}$ in the denominator. Thus, we expect pathologies for fields with spin $s>\frac{1}{2}$.

\subsection*{Case I: $s=\frac{1}{2}$}

Using the explicit expressions of $n(\bm p)$, $n(\bm p-\bm k)$, $\epsilon^{\lambda}_\mu(\bm k)$, and substituting into (\ref{projection correlator}) and (\ref{general final form of the EM correlator})  we obtain for the left-handed fields
\begin{eqnarray}
\nonumber
{\cal M}^{\lambda\,,\lambda'}(t_x,t_y)&=&\frac{1}{2}\alpha_{1/2}^2\delta^{\lambda\lambda'}\delta^{3}\left(\bm k+\bm k'\right)\int d^3 p~  e^{i\vartheta(\bm p, \bm k, t_x,t_y)} |\bm p|^2\sin^2\theta\\
&&\times \left(1+\frac{\lambda+\lambda'}{2}\cos\theta\right)\left(1+\frac{\lambda+\lambda'}{2}\frac{|\bm p|\cos \theta-k}{\sqrt{k^2+|\bm p|^2-2k|\bm p|\cos\theta}}\right)\,.
\label{massless LHF}
\end{eqnarray}
Similarly, the contribution from the right-handed fields is
\begin{eqnarray}
\nonumber
{\cal M}^{\lambda\,,\lambda'}(t_x,t_y)&=&\frac{1}{2}\alpha_{1/2}^2\delta^{\lambda\lambda'}\delta^{3}\left(\bm k+\bm k'\right)\int d^3 p~  e^{i\vartheta(\bm p, \bm k, t_x,t_y)} |\bm p|^2\sin^2\theta\\
&&\times \left(1-\frac{\lambda+\lambda'}{2}\cos\theta\right)\left(1-\frac{\lambda+\lambda'}{2}\frac{|\bm p|\cos \theta-k}{\sqrt{k^2+|\bm p|^2-2k|\bm p|\cos\theta}}\right)\,.
\label{massless RHF}
\end{eqnarray}
This exactly matches the results in \cite{Anber:2016yqr}, which were obtained using canonical quantization methods. 

\subsection*{Case II: $s \geq 1$}

In this case we find that there is a highly singular piece in the denominator that causes the integrand in (\ref{general final form of the EM correlator}) to blow up. Thus, we conclude that either the energy momentum tensor of general causal fields with spin $s\geq 1$ vanishes identically, or there cannot be a consistent description of these fields in flat space.  We could have anticipated this result since the amplitude for emitting or absorbing a massless particle with momentum $|\bm p|$ and spin $s$ vanishes like $|\bm p|^{s-\frac{1}{2}}$, as can be seen from (\ref{fields in terms of a and a dagger}). In fact, massless general causal fields cannot be used to build theories that describe long-range forces, as electromagnetism and gravity \cite{Weinberg:1965rz}. Instead, a massless spin $s$ particle  can be described by the symmetric potential $A^{\mu_1\mu_2...\mu_s}$, which transforms as $(\frac{s}{2},\frac{s}{2})$ under the Lorentz group. This is the direct generalization of the electromagnetic potential $A^\mu$ and the metric tensor $h^{\mu\nu}$ to higher spin fields. Such treatment has occupied a large portion of the literature on higher spin fields, and we refer the reader to \cite{Rahman:2015pzl} for an excellent review.

\subsection{Flat space energy-momentum tensor correlators: massive fields}

Despite the fact that massless fields are pathological in our formalism, massive fields are well behaved as we explain in this section. Let us define $\Pi^{\mu\nu}(p^0,\bm p)$ and $\bar \Pi^{\mu\nu}(p^0,\bm p)$ as
\begin{eqnarray}
\Pi^{\mu\nu}(p^0,\bm p)=p^{(\mu}t^{\nu) a_2...a_{2s}}p_{a_2}...p_{a_{2s}}\,, \quad \bar\Pi^{\mu\nu}(p^0,\bm p)=p^{(\mu}\bar t^{\nu) a_2...a_{2s}}p_{a_2}...p_{a_{2s}}\,.
\end{eqnarray}
Then, using (\ref{definition of energy-momentum correlator for the massive case}) we obtain
\begin{eqnarray}
\nonumber
\langle T^{\mu\nu}(\bm k, t_x)T^{\alpha\beta}(\bm k', t_y) \rangle_c&=&\frac{-i}{2}\alpha_s^2 m^{4-8s} \delta^3(\bm k+\bm k')\int \frac{d^3 p}{\omega(\bm p)\omega(\bm p-\bm k)}e^{i\vartheta (\bm p, \bm k, t_x,t_y)}\\
\nonumber
&& \mathbb{TR}\left\{\left[\begin{array}{cc}0 & \Pi^{\mu\nu}(p^0, \bm p) \\ \bar\Pi^{\mu\nu}(p^0, \bm p)&0 \end{array}\right] \left[\begin{array}{cc} m^{2s} \mathbb I & \Pi(p^0, \bm p) \\ \bar\Pi(p^0, \bm p)& m^{2s} \mathbb I  \end{array}\right]\right.\\
\nonumber
&&\times \left. \left[\begin{array}{cc} 0 & \Pi^{\alpha\beta}((p-k)^0, \bm p-\bm k) \\ \bar \Pi^{\alpha\beta}((p-k)^0, \bm p-\bm k)& 0 \end{array}\right]\right.\\
&&\times \left. \left[\begin{array}{cc} m^{2s} \mathbb I & \Pi((k-p)^0, \bm k-\bm p) \\ \bar\Pi((k-p)^0, \bm k-\bm p)& m^{2s} \mathbb I  \end{array}\right]\right\}\,.
\label{main expression for massive correlator in flat space}
\end{eqnarray}
The trace is over the Dirac indices in the representation $(s,0)\oplus(0,s)$. Expression (\ref{main expression for massive correlator in flat space}) is one of the main results in this work. One then can project (\ref{main expression for massive correlator in flat space})  along the polarization tensors to obtain the amplitude ${\cal M}^{\lambda\,\lambda'}$ as in (\ref{projection correlator}). Since both fields, the left and right handed, are present, we expect to have ${\cal M}^{++}={\cal M}^{--}$.

We can perform the matrix multiplication and take the Dirac traces of (\ref{main expression for massive correlator in flat space}) to find  
\begin{eqnarray}
\nonumber
\langle T^{\mu\nu}(\bm k, t_x)T^{\alpha\beta}(\bm k', t_y) \rangle_c&=&\frac{-i}{2}\alpha_s^2 m^{4-8s} \delta^3(\bm k+\bm k')\int \frac{d^3 p}{\omega(\bm p)\omega(\bm p-\bm k)}e^{i\vartheta (\bm p, \bm k, t_x,t_y)}\\
&&\times\left({\cal Z}_L^{\mu\nu;\alpha\beta}+{\cal Z}_R^{\mu\nu;\alpha\beta}\right)\,,
\label{energy momentum correlator for massive case with equal weight}
\end{eqnarray}
where
\begin{eqnarray}
\nonumber
{\cal Z}_L^{\mu\nu;\alpha\beta}&=&\Pi^{\mu\nu}(p^0, \bm p)\bar\Pi(p^0, \bm p) \Pi^{\alpha\beta}((p-k)^0, \bm p-\bm k)\bar \Pi((k-p)^0, \bm k-\bm p)\\
&&+m^{4s}\Pi^{\mu\nu}(p^0,\bm p)\bar \Pi^{\alpha\beta}(p^0-k^0,\bm p-\bm k)\,,\\
\label{ZLeft}
\nonumber
{\cal Z}_R^{\mu\nu;\alpha\beta}&=&\bar\Pi^{\mu\nu}(p^0, \bm p)\Pi(p^0, \bm p) \bar\Pi^{\alpha\beta}((p-k)^0, \bm p-\bm k)\Pi((k-p)^0, \bm k-\bm p)\\
&&+m^{4s}\bar\Pi^{\mu\nu}(p^0,\bm p) \Pi^{\alpha\beta}(p^0-k^0,\bm p-\bm k)
\label{ZRight}
\end{eqnarray}
In the following, we consider $s=\frac{1}{2}$ and $s=1$ as examples for the massive case. The computations for any $s\geq 2$ are extremely painful due to the cumbersome algebra of the $t$ matrices, and we feel that it is necessary to develop group theoretical tricks to deal with such cases. This task will be left for a future investigation.    

\subsection*{Case I: $s=\frac{1}{2}$}

The projection of the energy-momentum tensor correlator along the polarization eigenbases yields
\begin{eqnarray}
\nonumber
{\cal M}^{\lambda\,,\lambda'}(t_x,t_y)&=&\alpha_{1/2}^2\delta^{\lambda\lambda'}\delta^3(\bm k+\bm k')\int d^3p e^{i\vartheta (\bm p, \bm k, t_x,t_y)}\\
&& \times |\bm p|^2 \sin^2\theta\left[1+\frac{|\bm p|\cos\theta \left(|\bm p|\cos\theta -k\right)-m^2}{\sqrt{\left(m^2+|\bm p|^2\right)\left(m^2+|\bm p|^2+k^2 -2|\bm p|k\cos\theta\right)}} \right]\,.
\label{massive spin half}
\end{eqnarray}
In the $m=0$ limit,  it is trivial to see that (\ref{massive spin half}) is the sum of the contributions from left-handed (\ref{massless LHF}) and right-handed  (\ref{massless RHF}) fields. 

\subsection*{Case II: $s=1$}

In this case we find
\begin{eqnarray}
\nonumber
{\cal M}^{\lambda\,,\lambda'}(t_x,t_y)&=&\frac{\alpha_{1}^2}{2m^4}\delta^{\lambda\lambda'}\delta^3(\bm k+\bm k')\int d^3p e^{i\vartheta (\bm p, \bm k, t_x,t_y)}\frac{m^2 |\bm p|^2}{{\Gamma}}\sin^2\theta\\
\nonumber
 &&\times\left\{14 k^2|\bm p|^2+7m^2|\bm p|^2+3|\bm p|^4+8k^2\Gamma +4|\bm p|^2\Gamma\right.\\
\nonumber
&&\left.-k|\bm p|\cos\theta\left(8k^2+8m^2+19|\bm p|^2+16\Gamma\right) +|\bm p|^2\cos(2\theta)\left(10k^2+m^2+4|\bm p|^2+4\Gamma\right)\right.\\
&&\left. -5k |\bm p|^3\cos(3\theta)+|\bm p|^4\cos(4\theta)  \right\}\,,
\label{massive s 1 case}
\end{eqnarray}
where
\begin{eqnarray}
\Gamma=\sqrt{\left(m^2+|\bm p|^2\right)\left(m^2+|\bm p|^2+k^2-2k|\bm p|\cos\theta\right)}\,.
\end{eqnarray}

Now a few points are in order:
\begin{enumerate}
\item One should refrain from comparing the result (\ref{massive s 1 case}) with massive electrodynamics. First, the later theory is based on the representation $(\frac{1}{2}, \frac{1}{2})$ and uses the gauge potential $A_\mu$ instead of the physical fields. The $(\frac{1}{2}, \frac{1}{2})$ representation contains $s=1$ and $s=0$ spin fields and one needs to isolate the $s=1$ component by applying the constraint $\partial^\mu A_\mu=0$. This is in contrast with the $(1,0)\oplus(0,1)$ causal field construction\footnote{It was also shown in \cite{Birrell:1979jd} that the conformal anomalies of $(1,0)$ and $(\frac{1}{2},\frac{1}{2})$ fields are different.}. Second, one needs to bear in mind that the energy momentum tensor of massive electrodynamics (derived from Proca Lagrangian) is different from the one adopted in this work and given by (\ref{definition of energy momentum using the effective action}).  
\item It is easy to check that  $\langle T_-(k)T_-(-k) \rangle>0$ and $\langle T_+(k)T_+(-k) \rangle>0$ for both $s=\frac{1}{2}$ and $s=1$, and therefore, our theory respects unitarity in the checked cases. Eventhough one could have expected the $s=\frac{1}{2}$ theory to be unitary, the higher spin case is far from obvious.   
\item Since both fields $(0,1)$ and $(1,0)$ are present on equal footing, we have $\langle T_-(k)T_-(-k) \rangle=\langle T_+(k)T_+(-k) \rangle$. This is attributed to the fact that both the left (\ref{ZLeft}) and right (\ref{ZRight}) amplitudes enter the calculations of ${\cal M}^{\lambda\,,\lambda'}$ with the same weight. In principle, one could imagine a mechanism\footnote{For example, coupling to axions can lead to the preference of one component over the other \cite{Bartolo:2014hwa,Adshead:2015kza,Anber:2012du}. A similar effect can be obtained by turning on parity-violating operators.}that prefers the left over the right component, or vice versa. In this case the sum $\left({\cal Z}_L^{\mu\nu;\alpha\beta}+{\cal Z}_R^{\mu\nu;\alpha\beta}\right)$ in (\ref{energy momentum correlator for massive case with equal weight}) is replaced by the weighted sum $\left(W_L{\cal Z}_L^{\mu\nu;\alpha\beta}+W_R{\cal Z}_R^{\mu\nu;\alpha\beta}\right)$, where $W_{L,R}$ is the left(right) weight. Thus, we conclude that in general $\langle T_-(k)T_-(-k) \rangle\neq \langle T_+(k)T_+(-k) \rangle$. Such signal violates the macroscopic parity symmetry \cite{Lue:1998mq,Saito:2007kt}.
\end{enumerate}
%

\subsection{Energy-momentum tensor correlators in curved space: massless $s=\frac{1}{2}$ field}
\label{Energy-momentum correlators in curved space half case}

In the remaining of this section we calculate the energy-momentum tensor in FRW background for the massless $s=\frac{1}{2}$ field.  The energy-momentum tensor correlator is 
\begin{eqnarray}
\langle T^{\mu\nu}(x)T^{\alpha\beta}(y) \rangle=2i\langle {\cal O}^{\mu\nu}(x)G_0(x,y){\cal O}^{\alpha\beta}(y)G_0(y,x) \rangle\,,
\end{eqnarray}
where ${\cal O}^{\mu\nu}$ is given by (\ref{curved space vertex}). To compute this correlator, we first consider the quantity $D^a G^0$, which appears from the action of the vertex on $G^0$. Recalling that the Green's function is a bispinor with indices $G_{MN}^0$ such that $M,N=-1/2,1/2$, we have
\begin{eqnarray}
D^a G_{NM}^0=\eta^{ab}V_{b}^\rho \left[\partial_\rho\delta_N^P+\left(\Gamma_\rho\right)_N^P \right]G_{PM}^0\,.
\end{eqnarray} 
As we pointed out in Section (\ref{General causal fields in curved space}), the best strategy to compute this quantity in de Sitter background is to perform the Weyl transformation $\eta_{\mu\nu}\rightarrow \bar g_{\mu\nu}=\eta_{\mu\nu}\Omega^2$, $\Omega=\frac{1}{H\tau}$.  After some algebra we find
\begin{eqnarray}
\bar D^a G^0_{NM}=\Omega^{-1}\left[\partial^a\delta_N^P-\left(\Omega\right)^{-1}\left({\cal T}^{0a}\right)^P_N\Omega_{,0}\right]G_{PM}^0\,,
\end{eqnarray}
where ${\cal T}^{0a}=\sigma^a$ are the Lorentz generators of $s=1/2$ representation. Taking this into account, we obtain 
\begin{eqnarray}
\bar  {\cal O}^{ij}_{MN}G^0_{PN}=\frac{i}{2}\Omega^{-3}\left[\left(\sigma^i_{MN}\partial^j+\sigma_{MN}^j\partial^i\right)G^0_{PN}+\frac{\delta^{ij}}{\tau}G^0_{PM} \right]\,,
\end{eqnarray}
where we limited our computation to the spatial components of ${\cal O}^{ij}$ since these are the only components that survive after projecting the energy-momentum tensor along the helicity eigenstates. After straightforward algebra we find
\begin{eqnarray}
\nonumber
{\cal M}^{\lambda,\lambda'}(\tau_x,\tau_y)&=&2\alpha_{1/2}\Omega^{-3}(\tau_x)\Omega^{-3}(\tau_y)\delta^3(\bm k+\bm k')\int \frac{d^3 p}{(2\pi)^3}\bm p\cdot \bm \epsilon(\bm k) \bm p\cdot \bm\epsilon(-\bm k)n_c(\bm p)n_d(\bm k-\bm p)\\
\nonumber
&&\times\epsilon_i^\lambda(\bm k)\epsilon_l^\lambda(-\bm k)\mbox{tr}\left[\sigma^i\sigma^c\sigma^l\sigma^d \right]u^*(\bm p, \tau_x)u(\bm p, \tau_y)v^*(\bm p-\bm k, \tau_y)v(\bm p-\bm k, \tau_x)\,.
\end{eqnarray}
After taking the trace we finally obtain
\begin{eqnarray}
\nonumber
{\cal M}^{\lambda,\lambda'}(\tau_x,\tau_y)&=&2\alpha_{1/2}\Omega^{-3}(\tau_x)\Omega^{-3}(\tau_y)\delta^3(\bm k+\bm k')\int \frac{d^3 p}{(2\pi)^3}  |\bm p|^2\sin^2\theta \\
\nonumber
&&\times u^*(\bm p, \tau_x)u(\bm p, \tau_y)v^*(\bm p-\bm k, \tau_y)v(\bm p-\bm k, \tau_x)\\
&&\times \left(1-\frac{\lambda+\lambda'}{2}\cos\theta\right)\left(1-\frac{\lambda+\lambda'}{2}\frac{|\bm p|\cos \theta-k}{\sqrt{k^2+|\bm p|^2-2k|\bm p|\cos\theta}}\right)\,,
\end{eqnarray}
which exactly matches the result obtained previously in \cite{Anber:2016yqr} via canonical quantization methods. 

\section{Comments on the gravitational wave spectrum}
\label{Gravitational waves}

In the previous sections we determined the form of the energy-momentum tensor correlator, ${\cal M}^{\lambda\lambda'}$, for a general causal field with spin $s$. Given ${\cal M}^{\lambda\lambda'}$, we now can calculate the metric tensor correlator, and hence, the gravitational waves (GW) power spectrum. This is a standard procedure that can be found in various places in the literature, see e.g., \cite{Anber:2016yqr,Sorbo:2011rz}. However, we choose to review it here for the sake of completeness. To this end, we write the full metric as $g_{\mu\nu}$=$a^2(\tau)\left(\eta_{\mu\nu}+h_{\mu\nu}\right)$, where the background is taken to be FRW and $a(\tau)$ is the scale factor. After gauge fixing,  the linearized equations of motion for the tensor perturbations $h_{\mu\nu}$ are 
\begin{eqnarray}
\left( \partial_\tau^2+2{\cal H}\partial_\tau - {\nabla}^2\right)   h^{\mu\nu}(\bm x,\tau) = \frac{2}{m_p^2}~ T^{\mu\nu}(\bm x,\tau)\,,
\end{eqnarray}
where ${\cal H}=\frac{da}{a d\tau}$ is the conformal Hubble parameter and $m_p$ is the reduced Planck mass. One can also take the spatial Fourier transform of the above equation to obtain
\begin{eqnarray}
\left(\partial_\tau^2+2{\cal H}\partial_\tau +k^2\right)  h^{\mu\nu}(\bm k,\tau) = \frac{2}{m_p^2}~ T^{\mu\nu} (\bm k, \tau)\,.
\end{eqnarray}
The gravitational perturbations $h^{\mu\nu}$ can be expressed in terms of the two circular helicity modes $h_{\pm}$ as
\begin{eqnarray}
\label{h circular}
h^{\mu\nu} (\bm k, \tau) = \sum_{\lambda = \pm} \epsilon_{\lambda}^{\mu}(\bm k) \epsilon_{\lambda}^{\nu}(\bm k) h_{\lambda} (\bm k,\tau) \,.
\end{eqnarray}
This decomposition  is consistent with the transverse and traceless conditions $h_{00}=h_{0i}=h_i^i=h_{i,j}^i=0$, and hence, it projects only the physical degrees of freedom. One then uses the projection operator $\Pi_{\mu\nu,\alpha\beta} (\bm k) = {\cal P}_{\mu\alpha} (\bm k) {\cal P}_{\nu\beta} (\bm k) - \frac{1}{2}{\cal P}_{\mu\nu} (\bm k) {\cal P}_{\alpha\beta} (\bm k)$ along with the identity $\epsilon_{\lambda}^{\mu\nu}(\bm k) ~\Pi_{\mu\nu}^{~~~\alpha\beta} (\bm k) = \epsilon_{\lambda}^{\alpha\beta}(\bm k)$ to project out the physical components of the tensor perturbations $h_{\lambda=\pm}$. Finally, we find that the equation of motion of  $h_{\lambda=\pm}$ reads
\begin{eqnarray}
\left(\partial_\tau^2+2{\cal H}\partial_\tau +k^2 \right) h_{\lambda}(\bm k,\tau) =  \frac{1}{m_p^2}~ \epsilon_{-\lambda}^{\mu\nu}(\bm k)  ~ T_{\mu\nu} (\bm k, \tau)\,.
\label{main h equation}
\end{eqnarray}
The solution of (\ref{main h equation}) is given by 
\begin{eqnarray}
h_\lambda(\bm k,\tau) =h_\lambda^{\mbox{\scriptsize hom}}(\bm k,\tau)+ \frac{2}{m_p^2} \int d\tau' G_{k}(\tau,\tau') ~\epsilon_{-\lambda}^{\mu\nu}(\bm k) ~ T_{\mu\nu} (\bm k, \tau')\,,
\end{eqnarray}
where $h_\lambda^{\mbox{\scriptsize hom}}(\bm k,\tau)$ is the homogeneous solution of (\ref{main h equation}) and $ G_{k}(\tau,\tau')$  is the retarded Green's function of the differential operator on the left hand side of (\ref{main h equation}), i.e.,
\begin{eqnarray}
\left( \partial_\tau^2+2{\cal H}\partial_\tau  +k^2  \right)G_k(\tau,\tau')=\delta(\tau-\tau')\,.
\end{eqnarray}
Finally, with the aid of definition (\ref{projection correlator}), the correlator $\langle h_{\lambda}(\bm k,\tau)h_{\lambda'}(\bm k,\tau) \rangle$ is given by the expression:
\begin{eqnarray}
\langle h_\lambda(\bm k,\tau) h_{\lambda'}(\bm k',\tau) \rangle &=&-i  \frac{4}{m_p^4}\delta(\bm k-\bm k') \int d\tau' \tau''  G_{k}(\tau,\tau') G_{k'}(\tau,\tau'') ~{\cal M}^{\lambda \lambda'} (\tau',\tau'') \,.\label{h correlator} 
\end{eqnarray}

Expression (\ref{h correlator}) gives the GW spectrum in terms of a complicated function ${\cal M}^{\lambda \lambda'} (\tau',\tau'')$. Irrespective of these complications, the conformal symmetry of de Sitter space is sufficient to determine the GW correlators. For a Hubble parameter much larger than the particle mass, $H\gg m$, the general form of the GW correlators take the form
\begin{eqnarray}
\nonumber
\langle h_-(\bm k,\tau) h_{-}(\bm k,\tau) \rangle &\cong&\frac{H^2}{\pi^2 m_p^2}\left[1+{\cal N}_1\frac{H^2}{m_p^2}\left(\frac{k}{k_0}\right)^{n_{-}}    \right]\,,\\
\langle h_+(\bm k,\tau) h_{+}(\bm k,\tau) \rangle&\cong& \frac{H^2}{\pi^2 m_p^2}\left[1+{\cal N}_2\frac{H^2}{m_p^2} \left(\frac{k}{k_0}\right)^{n_{+}}    \right]\,,
\label{}
\end{eqnarray}
where the first term is the vacuum contribution (the homogeneous term), ${\cal N}_{1(2)}$ are left(right) coefficients, $k_0$ is a reference wavevector, and $n_{\pm}$ are exponents that quantify the deviation from an exact scale invariance. In a quasi de Sitter space we expect to have $|n_{\pm}|\ll 1$. In addition,  any imbalance between the production of $(s,0)$ and $(0,s)$ fields translates into ${\cal N}_{1}\neq {\cal N}_{2}$. Therefore, an observation of parity violation  in the primordial GW power spectrum, i.e., $\langle h_-(\bm k,\tau) h_{-}(\bm k,\tau)\neq \langle h_+(\bm k,\tau) h_{+}(\bm k,\tau) \rangle $,  will point to parity violating physics in the early Universe.  The coefficients ${\cal N}_{1,2}$ are proportional to $\alpha_{s}$ that appears in the vertex operator (\ref{O for massive fields}). Although in this work we do not give a detailed particle physics model that explains the origin of $\alpha_{s}$, we expect on general grounds that a sufficient mechanism can produce higher spin bosonic fields from vacuum  with an effective $\alpha_{s}\gg1$. On the other hand, the production of fermionic higher spin fields from vacuum yields $\alpha_{s}\sim 1$, which is a result of the Pauli exclusion principle.

\section{Discussion}
\label{Discussion}

Higher spin fields remains one of the most challenging topics in physics. In particular,  their coupling to gravity or electromagnetism has proven to be a highly non-trivial task. Over the past few years there has been a noticeable progress toward a better understanding of higher spin fields, their interactions, as well as their behavior in external background fields, see \cite{Boulanger:2006gr,Porrati:2008ha,Didenko:2014dwa,Gomez-Avila:2013qaa,Buchbinder:2011xw,Reshetnyak:2012ec,Henneaux:2013gba,Cortese:2013lda} for a non-comprehensive list of references.  In this work we argued that irrespective of these difficulties, general principles of Poincar\'e covrainace and analaticity of the Green's functions are enough to dictate the form of the energy-momentum tensor correlators and to conclude that the production of higher spin fields from vacuum is accompanied by large-scale signals that can be printed  in the Cosmic Microwave Background (CMB) or detected in terrestrial or space-based gravitational wave interferometry (GWI) \cite{Cook:2011hg}.

The explicit form of the energy-momentum tensor correlator depends on the representation used to describe the higher spin particles. There are two main methods in the literature that are used to describe fields with spin $s\geq \frac{1}{2}$: the Fronsdal construction, which is based on the $(s,s)$ symmetric representation, and the general causal (Weinberg-type) fields that transform under $(s,0)$, $(0,s)$, or $(s,0)\oplus(0,s)$. In this work we chose to use the latter formalism as it provides a short-cut way to make our points explicit and avoid the hassle of introducing spurious degrees of freedom and/or auxiliary fields. Weinberg-type fields are physical, and therefore, no question of inconsistencies or ghosts can appear. The lack of a Lagrangian formulation of these fields forced us to postulate an energy-momentum tensor. Its form is inspired by the transformation properties under the Poincar\'e group. 

The fact that we can infer most information about the energy-momentum tensor correlators from general considerations of Poincar\'e group and analytic structure of Green's function may sound puzzling. Here, we provide an argument that elucidates this observation.   The traditional method of computing field theory correlators in cosmology is to start from the Bunch-Davies vacuum $|0\rangle_{a}$ and then evolve the field theory operators using the Bogoliubov transformation. Let $\{a_i,a_i^\dagger\}$ be a set of annihilation and creation operators, such that the operators $\{a_i\}$ annihilate the vacuum: $a_i|0\rangle_{a}=0$. In the presence of a time varying background the set $\{a_i,a_i^\dagger\}$ evolves into a new set of operators $\{b_i,b_i^\dagger\}$. The latter ones are linear combinations of $\{a_i,a_i^\dagger\}$. Therefore, an energy-momentum tensor correlator of the form $\langle 0|\sum_{ij} \left({\cal C}_i a_i+{\cal C}_i^* a_i^\dagger\right)\left({\cal C}_j a_j+{\cal C}_j^* a_j^\dagger\right) |0\rangle_a$, which is vanishing in the Bunch-Davies vacuum,  will evolve into  the correlator  $\alpha\langle 0|\sum_{ij} \left({\cal C}_i b_i+{\cal C}_i^* b_i^\dagger\right)\left({\cal C}_j b_j+{\cal C}_j^* b_j^\dagger\right) |0\rangle_a$, which is not zero. Now, one can immediately see that the final product is a collection of correlators of the form $\langle 0| b_i^\dagger b_j|0\rangle_a$, which are nothing but Green's functions of on-shell particles. Therefore, all information about the energy-momentum tensor correlators are encoded in the analytic structure of Green's functions. The coefficients $\{{\cal C}_i\}$ are group theoretical constants that depend on the specific representation at hand, while $\alpha$ is some number that depends on the specifics of the physical process that lead to the particle production, e.g., parametric resonance. This is ${\cal O} (1)$ number for fermions, thanks to the Pauli-blocking. However, one can have $\alpha \gg1$ for bosonic fields. Thus, one expects that an abundant production of higher spin bosons can leave an ``in principle detectable" imprint on the CMB or GWI.         

Another question concerns the parity symmetry of the Universe. We have shown in this work that any mechanism that causes an imbalance between the on-shell production of $(s,0)$ and $(0,s)$ fields leads to imbalance between the left- and right-handed energy momentum correlators, $\langle T_-(k)T_-(-k) \rangle\neq \langle T_+(k)T_+(-k) \rangle$, and hence, a violation of the macroscopic parity symmetry. This generalizes the previous observation that the production of fermions with a definite helicity is accompanied by the generation of chiral gravitational waves \cite{Anber:2016yqr}. 

General causal fields is useful as a proof of concept for a few ideas that we discussed in this work. However, we found that this formalism is not suitable to study the production of massless fields in flat space. In addition, one needs to deal with  cumbersome algebra  for $s>1$. There are also difficulties regarding the ordering of the covariant derivatives in curved background. All these questions call for more investigations of the production of higher spin fields in a cosmological context, either using the general causal field recipe or by other means.

\acknowledgments
I would like to thank  Dario Francia, Elias Kiritsis, Erich Poppitz, Rakibur Rahman, Riccardo Rattazzi, Augusto Sagnotti, and Sergey Sibiryakov for useful discussions, and Lorenzo Sorbo for critical comments on the manuscript. Special thanks go to Eray Sabancilar for many intense conversations that inspired the idea of this work and for collaboration at various stages of this project. Also, I would like to thank Crete Center for Theoretical Physics for the hospitality and the Dean's Office at Lewis \& Clark College for the faculty retreat at the Pacific ocean coast,  where parts of this work were completed. My research is supported by NSF grant PHY-1720135 and the Murdock Charitable Trust.

\bibliography{references_messengers_higher_spins}

\bibliographystyle{JHEP}

\end{document}